\documentclass[acmlarge]{acmart}
\usepackage{tikz}
\usepackage{amsmath}
\usepackage{float}
\usepackage{algorithm}
\usepackage{algorithmic}
\usepackage{subcaption}
\usepackage{enumitem}
\usepackage{appendix}
\usepackage{multirow}
\usepackage{booktabs}
\usepackage{siunitx}
\usepackage{colortbl}
\usepackage{xcolor}

\AtBeginDocument{%
  }

\copyrightyear{2024}
\acmYear{2024}
\acmMonth{12}
\acmArticle{172}
\acmVolume{8}
\acmNumber{4}
\acmJournal{IMWUT}

\setcopyright{acmlicensed}
\copyrightyear{2024}
\acmYear{2024}
\acmDOI{https://doi.org/10.1145/3699772}

\begin{document}

\title{mmSpyVR: Exploiting mmWave Radar for Penetrating Obstacles to Uncover Privacy Vulnerability of Virtual Reality}

\author{Luoyu Mei}
\orcid{0000-0003-2338-0256}
\affiliation{
  \institution{Southeast University}
  \city{Nanjing}
  \country{China}
  \institution{City University of Hong Kong}
  \city{Hong Kong}
  \country{China}
}
\email{lymei-@seu.edu.cn}

\author{Ruofeng Liu}
\orcid{0000-0001-9804-3727}
\affiliation{
  \institution{Michigan State University}
  \city{East Lansing}
  \country{United States}
}
\email{liuruofe@msu.edu}

\author{Zhimeng Yin}
\orcid{0000-0002-3548-3335}
\affiliation{
  \institution{City University of Hong Kong}
  \city{Hong Kong}
  \country{China}
}
\email{zhimeyin@cityu.edu.hk}

\author{Qingchuan Zhao}
\orcid{0000-0003-0163-2846}
\affiliation{
  \institution{City University of Hong Kong}
  \city{Hong Kong}
  \country{China}
}
\email{cs.qczhao@cityu.edu.hk}

\author{Wenchao Jiang}
\orcid{0000-0001-6765-2466}
\affiliation{
  \institution{Singapore University of Technology and Design}
  \city{Singapore}
  \country{Singapore}
}
\email{wenchao\_jiang@sutd.edu.sg}

\author{Shuai Wang}
\orcid{0000-0002-3609-2205}
\affiliation{
  \institution{Southeast University}
  \city{Nanjing}
  \country{China}
}
\email{shuaiwang@seu.edu.cn}

\author{Shuai Wang*}
\orcid{0000-0003-2766-1135}
\affiliation{
  \institution{Southeast University}
  \city{Nanjing}
  \country{China}
}
\email{shuaiwang\_iot@seu.edu.cn}

\author{Kangjie Lu}
\orcid{0000-0002-4763-7354}
\affiliation{
  \institution{University of Minnesota}
  \city{Minneapolis}
  \country{United States}
}
\email{kjlu@umn.edu}

\author{Tian He}
\orcid{0000-0001-6062-2619}
\affiliation{
  \institution{Southeast University}
  \city{Nanjing}
  \country{China}
}
\email{tianhe@seu.edu.cn}

\renewcommand{\thefootnote}{}
\footnotetext{Shuai Wang* (shuaiwang\_iot@seu.edu.cn) is the corresponding author.}
\renewcommand{\thefootnote}{\arabic{footnote}}

\renewcommand{\shortauthors}{Mei et al.}

\begin{abstract}
    Virtual reality (VR), while enhancing user experiences, introduces significant privacy risks. This paper reveals a novel vulnerability in VR systems that allows attackers to capture VR privacy through obstacles utilizing millimeter-wave (mmWave) signals without physical intrusion and virtual connection with the VR devices. We propose mmSpyVR, a novel attack on VR user's privacy via mmWave radar. The mmSpyVR framework encompasses two main parts: (i) A transfer learning-based feature extraction model to achieve VR feature extraction from mmWave signal. (ii) An attention-based VR privacy spying module to spy VR privacy information from the extracted feature. The mmSpyVR demonstrates the capability to extract critical VR privacy from the mmWave signals that have penetrated through obstacles. We evaluate mmSpyVR through IRB-approved user studies. Across 22 participants engaged in four experimental scenes utilizing VR devices from three different manufacturers, our system achieves an application recognition accuracy of 98.5\% and keystroke recognition accuracy of 92.6\%. This newly discovered vulnerability has implications across various domains, such as cybersecurity, privacy protection, and VR technology development. We also engage with VR manufacturer Meta to discuss and explore potential mitigation strategies. Data and code are publicly available for scrutiny and research.~\footnote{\url{https://github.com/luoyumei1-a/mmSpyVR/}}
\end{abstract}



\maketitle


\section{Introduction}
\label{sec1_introduction}
Virtual reality (VR) gains widespread popularity among enthusiasts and professionals in various activities, ranging from immersive gaming experiences to virtual chatting and online shopping~\cite{ZGaming, ThingShare}. These users interact with their VR devices through VR actions, such as body and hand controller motion. However, these VR actions inadvertently reveal the user’s privacy, specifically the user’s activity type and keystroke typing.

Existing methods for compromising the privacy of VR users fall into two distinct categories: (i) Physically entering the VR user’s environment. (ii) Virtually establishing a connection with the VR devices. In the first category, researchers hack into webcams and place hidden cameras~\cite{10.1145/3580779, 10.1145/3550325} to record VR actions~\cite{HoloLogger, Keylogging}. These approaches face constraints due to security barriers and non-line-of-sight scenarios. In the second category, researchers attempt to directly access the internal sensors by hacking the devices~\cite{Face-Mic, USENIX24}. The viability of these methods is limited due to the requirement of establishing a connection with the victim device and the fact that the user’s hand movements in the virtual scene do not reflect their actual movements. 

Recent studies explore the utilization of wireless signals~\cite{ding2020rfnet}, such as WiFi and mmWave, for VR privacy spying~\cite{Privacy_Leakage2, ijcai2024p131}. However, WiFi-based methods are susceptible to interference from environmental signals, which affects detection accuracy~\cite{10152700, SALIM2024102074, WiCAR, Multi_Adversarial, Mobi_Track}. Moreover, they struggle to achieve the high-precision hand position detection and reconstruction necessary for effective VR privacy spying. Meanwhile, existing mmWave-based posture recognition methods~\cite{mmASL, Wall_Matters} are not directly applicable to VR privacy spying, as these approaches fail to extract VR privacy information from mmWave signals attenuated by obstacles. The challenge lies in the complex relationship between VR user motions and their corresponding input. For instance, if a VR user clicks with an identical click twice but changes the headset orientation, it results in different inputs. Subsequently, existing mmWave-based motion sensing research does not explore the utilization of VR privacy spying.

In contrast to the aforementioned approaches, this paper proposes a novel side-channel vulnerability that exploits mmWave radar to penetrate obstacles \textbf{\textit{to spy on VR user's privacy without physical intrusion and virtual connection with the VR devices.}} We identify several opportunities to exploit this vulnerability. Firstly, the substantial bandwidth and high frequency provided by mmWave signals offer precision enhancement, improving the accuracy of sensory data~\cite{End-to-End}. Moreover, the penetration capabilities of mmWave signals pose significant threats to the privacy and security of VR devices~\cite{mircoDoppler}. In addition, the unique reflection characteristics of VR controllers and headsets present an opportunity for motion tracking, enhancing the point cloud features and reducing point cloud sparsity. Our system utilizes these opportunities to uncover the user's privacy.

To intercept VR user activity from mmWave signals through obstacles, we face three unique challenges: (i) What impact do indoor obstacles have on millimeter wave signals for VR users? (ii) How to precisely reconstruct user actions from sparse point clouds across obstacles? Due to the sparsity of mmWave point clouds, single-frame actions are insufficient to discern whether a user is inputting characters and words. (iii) How to establish a correspondence between VR actions and the user's VR privacy information? The same actions at different times correspond to different VR privacy information, requiring contextual information for accurate interpretation.

To address these challenges, the proposed \textbf{mmSpyVR} system incorporates a transfer learning-based \textbf{VR feature extraction} module and an attention-based \textbf{VR privacy spying} module. We develop a novel data augmentation technique to mitigate the impact of signal attenuation caused by obstacles, enhancing the system’s robustness in real-world environments. The privacy spying module leverages augmented point clouds of the user's body for comprehensive activity surveillance, focusing primarily on keyboard input activities. By recognizing VR motion, keyboard layout, position, and key presses, the system monitors various activity types and keystroke inputs across diverse VR environments. A continual learning approach is used to adapt to new activities, while a network with contextual temporal understanding captures VR privacy information embedded in user actions. The model further integrates multi-task learning for key press detection and keystroke prediction, accounting for different orientations, distances, and head movements during input processes. To the best of our knowledge, this is the first system to exploit mmWave signals for spying on VR users' privacy through obstacles, revealing a new security vulnerability. Our key contributions are as follows:
\begin{itemize}[topsep=0pt]
    \item To the best of our knowledge, mmSpyVR is the first attack on VR users' privacy via mmWave radar without physical and virtual connection with VR users. We investigate the feasibility of VR spying in various practical system settings and verify the ability of mmWave signals to penetrate obstacles.
    \item Technically, our design incorporates a VR feature extraction module founded on transfer learning to extract VR features despite the presence of obstacles. Additionally, we design an attention-based VR privacy spying model aimed at spying on the private information of VR users.
    \item Experimentally, we evaluate our system on 22 participants in 4 experimental scenes utilizing commodity VR devices from three different manufacturers and collect 12TB data in total. Experimental results show that mmSpyVR achieves 98.5\% and 92.6\% accuracy in activity type and keystroke spying.
\end{itemize}

\section{Background and Motivation}
\label{sec2_motivation}
In this section, we first discuss the privacy leakage issues in VR devices and the potential gains for attackers exploiting these vulnerabilities. Following that, we introduce the fundamental principles of mmWave sensing and analyze its feasibility for VR privacy spying. We demonstrate that mmWave signals can be exploited to spy on VR users, as their actions inadvertently leak sensitive information, even in the presence of obstacles.

\subsection{Privacy Leakage and Attacker Gain}
VR devices, while enhancing user experiences, also present significant privacy risks. Research indicates that users' motions during VR interactions leak sensitive data~\cite{Privacy_Leakage1, Privacy_Leakage2, Privacy_Leakage3}. Attackers exploit these motions to identify activities and keystroke inputs, potentially accessing critical information like banking credentials, passwords, and private messages~\cite{10.1145/3503161.3548386, 9382914, De-anonymization, Valluripally_Modeling, Valluripally_Detection, Radio2Text}. Body movements reveal activity types, while hand movements, particularly during virtual keyboard use, expose keystroke details~\cite{ZGaming, VRChat23, 10.1145/3580861}. This vulnerability is especially concerning as attackers are able to utilize mmWave technology to capture these motions through obstacles, unbeknownst to users~\cite{HiddenReality, mmPhone, MILLIEAR}. The penetrative ability of mmWave signals makes this threat model both realistic and feasible. A detailed analysis of privacy risks is provided in the discussion in Section~\ref{sec7_discussion}.

\subsection{Principles of mmWave Sensing}
This subsection describes the principle of mmWave radar sensing, which is crucial for obtaining the VR user's mmWave point cloud, a pivotal step in interpreting the spatial context of VR actions~\cite{DomainIndependent, IndexPen, Orientation-Aware, RadarNet}. The radar transmits and receives reflected signals, carrying essential data about the user's VR actions~\cite{mmWaveSurvey1, mmWaveSurvey2}. Range measurement is achieved by estimating the Time of Flight (ToF) of the received signal, yielding a complex intermediate frequency (IF) signal: $S^M(t) =  e^{j2\pi(\frac{B\tau}{T} t + f_c\tau)}$, where $\tau$ represents the ToF. The range $d$ is then calculated as $d = \frac{\tau c}{2}$, where $c$ is the speed of light. Velocity measurement utilizes phase differences of the IF signal between chirps. When a target moves at velocity $v_t$, the phase change $\Delta\phi$ between chirps allows us to calculate the velocity as $v_t=\frac{\lambda \Delta \phi}{4\pi T}$, where $\lambda$ is the wavelength and $T$ is the chirp duration. Angle measurement is based on phase differences of the IF signal across antennas. The angle of arrival $\theta$ is calculated using $\theta = cos^{-1}(\frac{\lambda\Delta\phi}{2\pi l})$, where $l$ is the distance between antennas and $\Delta\phi$ is the phase difference. These measurements collectively enable the creation of a detailed mmWave Doppler and point cloud representation of the VR user's actions.

\subsection{Penetration and Privacy Spying}

\begin{figure}[h]
    \centering
    \includegraphics[width=0.5\linewidth]{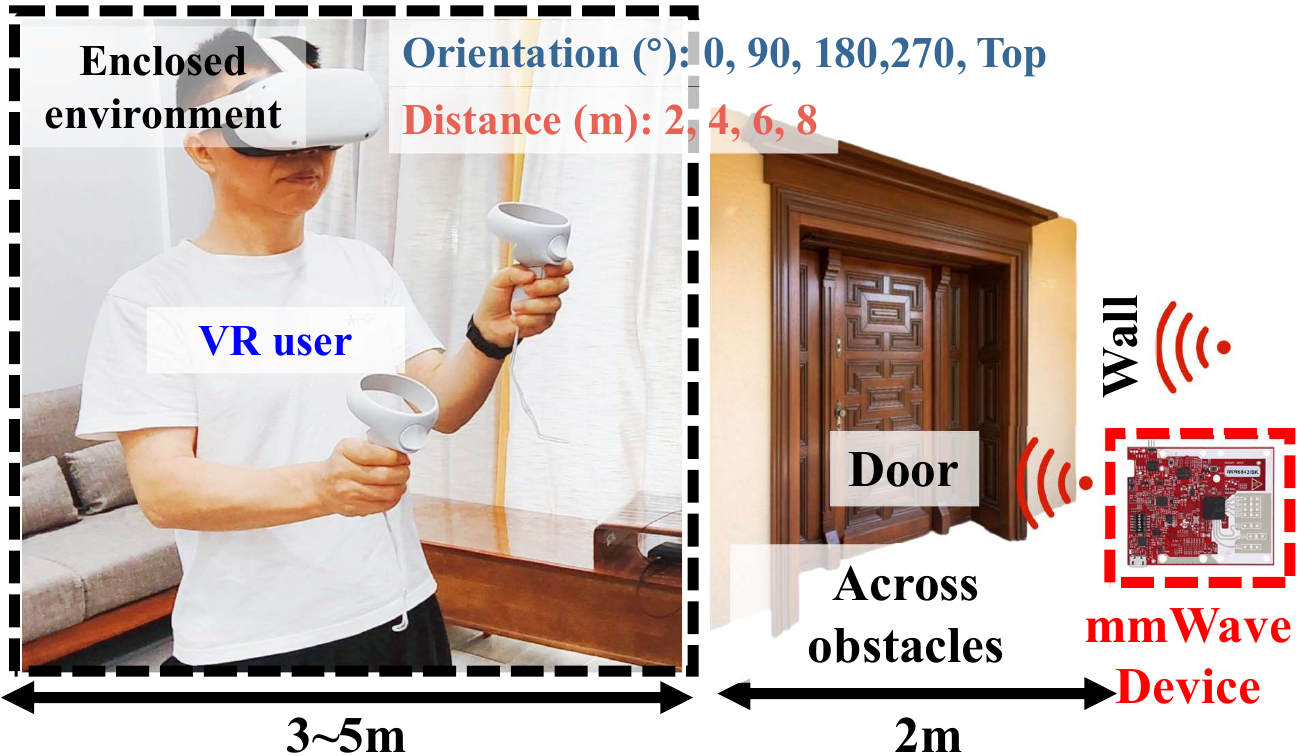}
    \caption{Privacy spying scenario.}
    \label{fig:scenario}
\end{figure}

We conduct two real-world experiments in the scenarios depicted in Fig.~\ref{fig:scenario} to further motivate the spy via mmWave radar and answer the following two critical research questions (RQs):

\begin{itemize}
    \item \textbf{RQ1:} What impact do indoor obstacles have on millimeter wave signals for VR users?
    \item \textbf{RQ2:} How to reconstruct user motion from sparse point clouds generated from mmWave signals?
\end{itemize}

We consider prevalent privacy spying scenarios~\cite{USENIX24, Privacy_Leakage2} where a VR user engages with VR in private and public spaces. To investigate mmWave sensing potential for VR privacy spying through obstacles, we conduct motivation experiments involving a user moving within a 10-meter room. We conduct sensing from various angles and distances, ranging from 2 to 8 meters and covering angles from $0^\circ$ to $360^\circ$, including a top view perspective from the rooftop. This setup aims to test mmWave penetration capabilities in real-life conditions without direct line-of-sight, while the VR user remains unaware of the non-line-of-sight sensing device. We assume researchers  \textit{cannot} compromise the VR device's hardware and software. Our goal is to determine how effectively mmWave radar recognize VR user actions and gestures despite physical obstacles.

The first experiment shows that mmWave signals are capable of penetrating obstacles, such as wooden doors and brick walls~\cite{Wall_Matters}. This allows the attackers to perform sensing without physical and virtual access to the VR devices. The second experiment shows that mmWave signals carry the motion information of the VR user~\cite{End-to-End}, such as their body and hand movements, which reflect the VR privacy.

\subsubsection{Penetration Experiment} 
We examine how mmWave signals penetrate obstacles and detect the VR user's motion behind them. We conduct an experiment where the attacker's mmWave device is located behind obstacles about four meters away from the VR user, who is waving the controller to play a game, facing the user's front. The mmWave radar emits signals that travel through the obstacles and reflect from the user's body. We utilize the reflected signals to generate range-Doppler maps~\cite{Vid2Doppler} that plot the distance and velocity of the detected user.

\begin{figure}[h]
\centering
    \begin{subfigure}[t]{0.24\linewidth}
        \includegraphics[width=\linewidth]{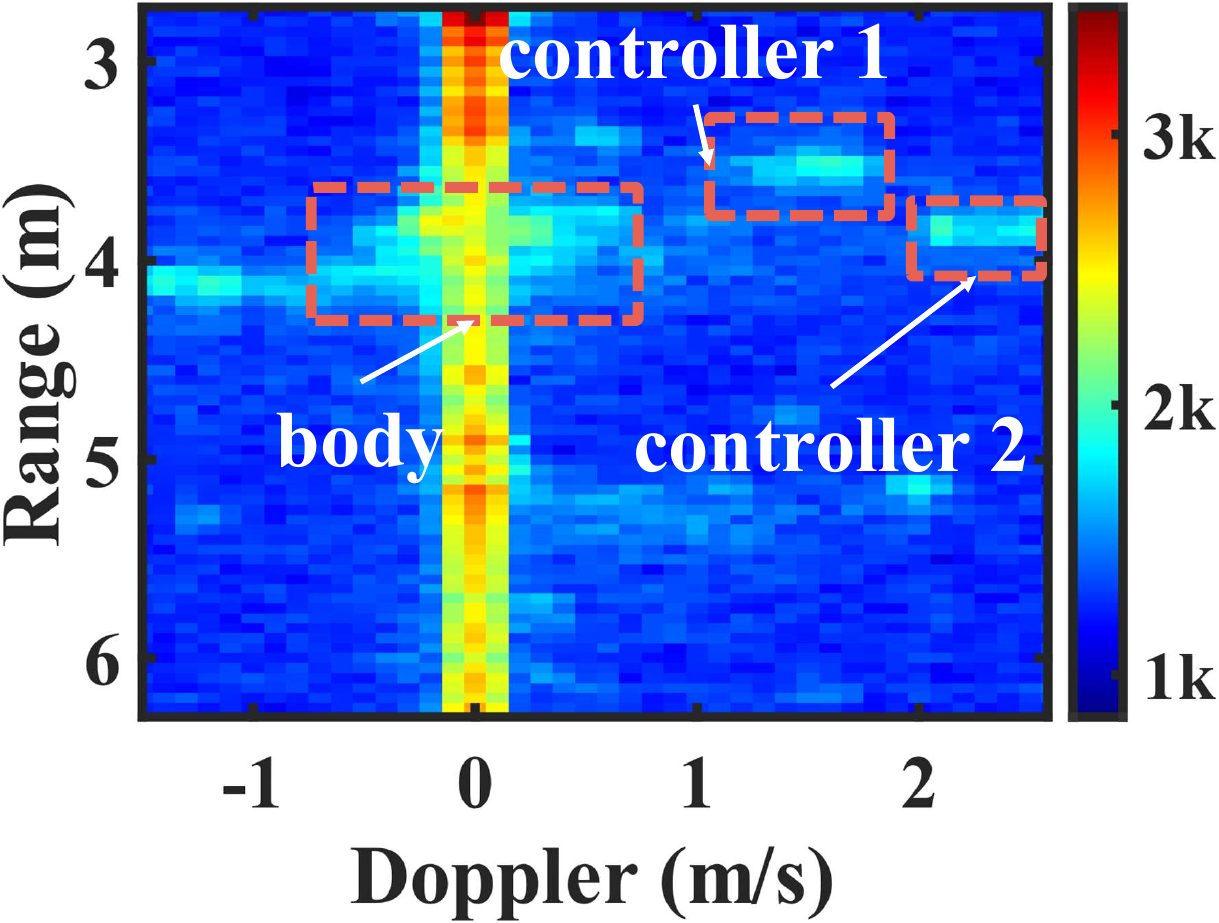}
        \caption{Unobstructed.}
        \label{fig:rdunobstructed}
    \end{subfigure}
    \begin{subfigure}[t]{0.24\linewidth}
        \includegraphics[width=\linewidth]{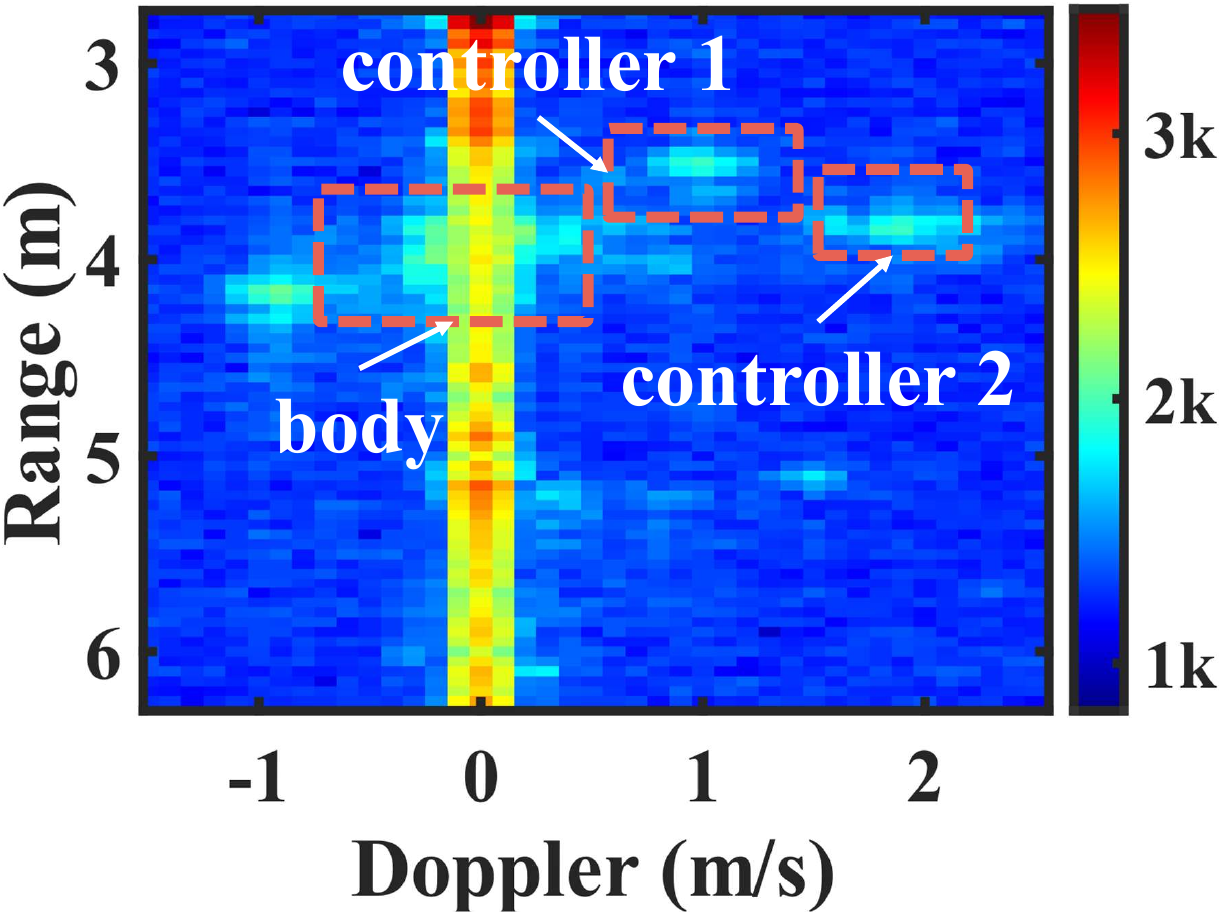}
        \caption{Wooden door.}
        \label{fig:rdwood}
    \end{subfigure}
    \begin{subfigure}[t]{0.24\linewidth}
        \includegraphics[width=\linewidth]{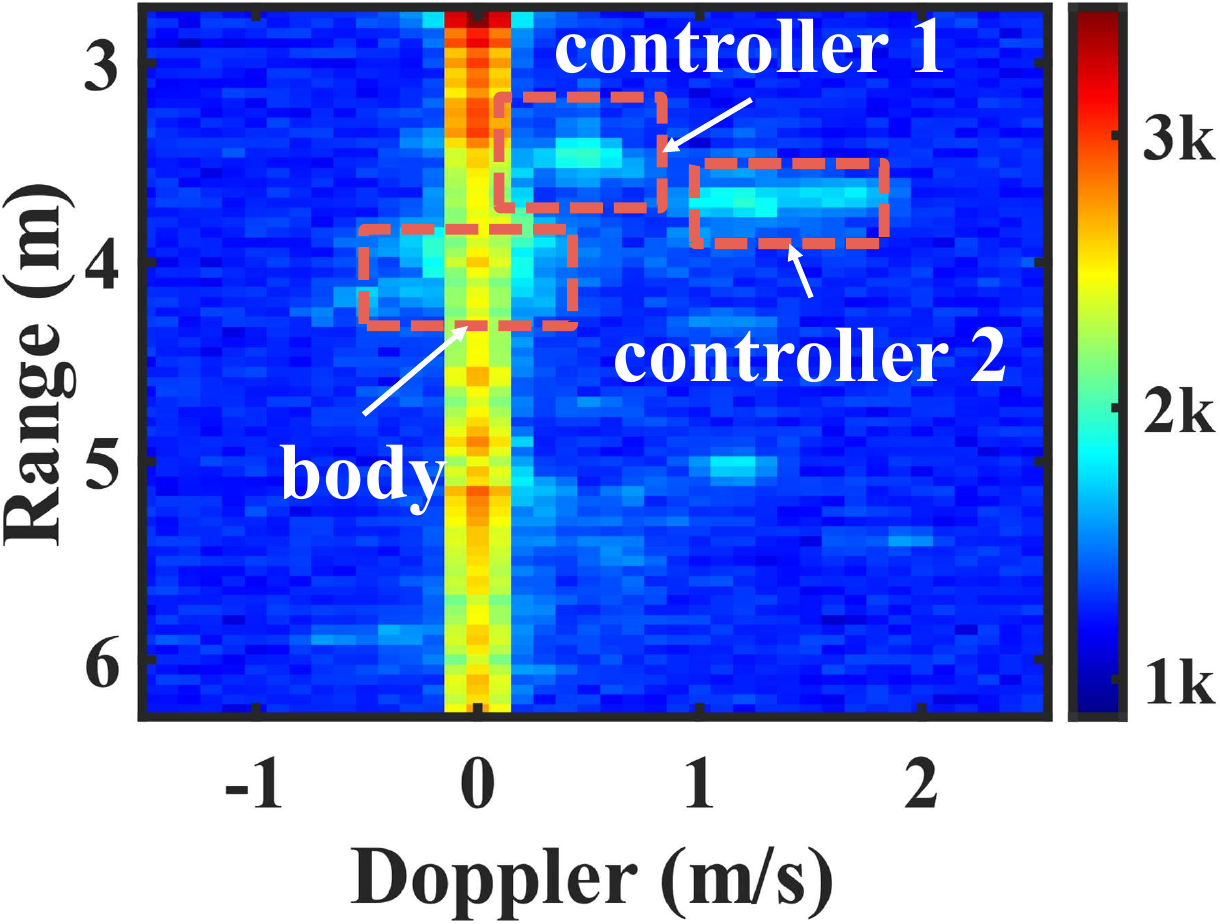}
        \caption{Brick wall.}
        \label{fig:rdbrick}
    \end{subfigure}
    \begin{subfigure}[t]{0.24\linewidth}
        \includegraphics[width=\linewidth]{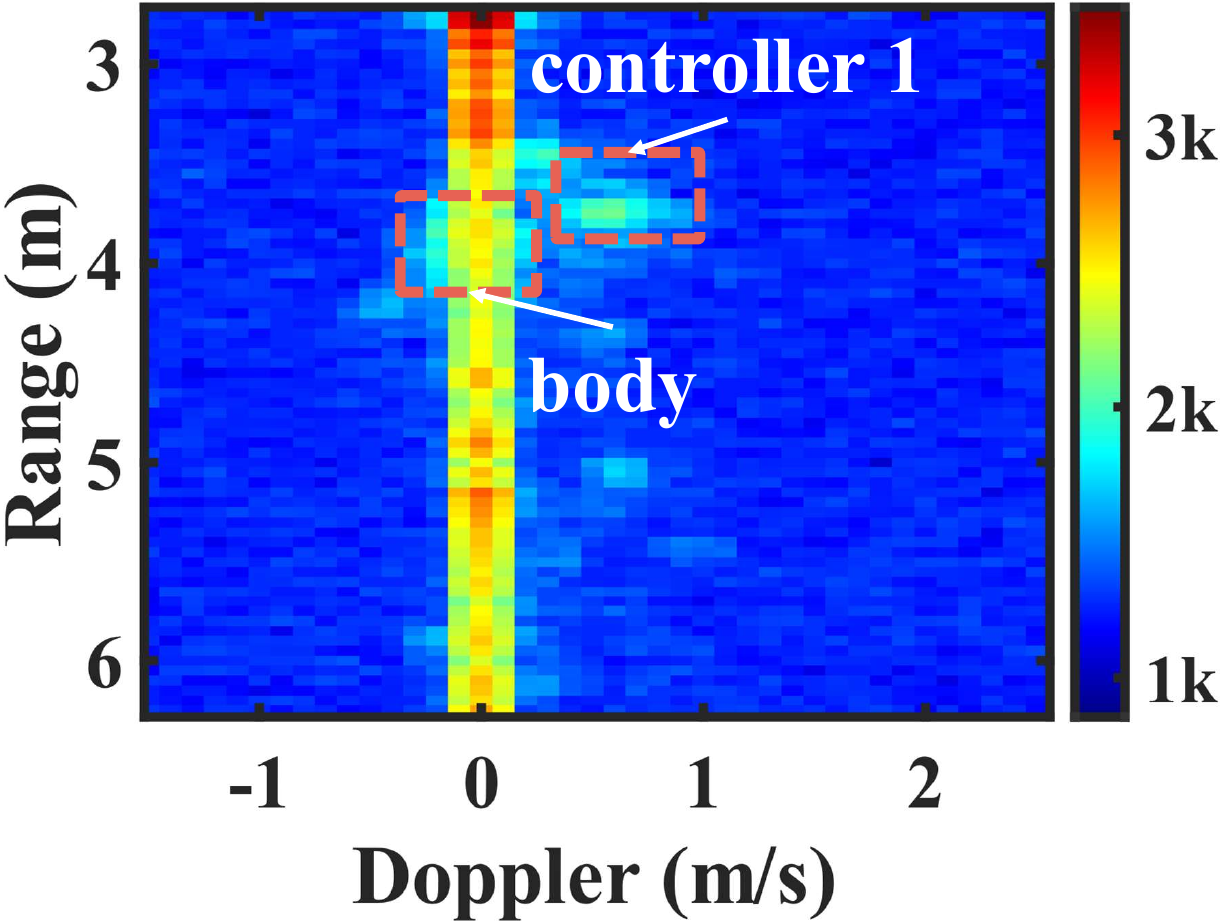}
        \caption{Wooden door \& brick wall.}
        \label{fig:rdwoodbrick}
    \end{subfigure}\vspace{-1em}
\caption{Range-Doppler maps of the mmWave signal after penetrating obstacles.}
\label{fig:penetration}
\end{figure}

Fig.~\ref{fig:penetration} illustrates the VR user's range-Doppler map when pressing keyboard buttons, with the horizontal axis as the distance from the device (Range) and the vertical axis as the relative speed to the device (Doppler). The yellow highlighted areas in the maps correspond to the parts of the body, controller 1 and controller 2. The signal strength diminishes with increased obstacle thickness, i.e., the decreased amount of yellow highlighted areas on the maps. However, even with both the wooden and brick obstacles, the mmWave radars retain enough resolution to precisely recognize the user body and the VR controller.

\subsubsection{VR Action Tracking Experiment} 
After penetrating obstacles, this experiment aims to evaluate whether the reflected mmWave signal contains sufficient VR action information and to qualify the amount of data extractable from such signals. We achieve this by extracting features from the reflected signals and focusing on the point clouds associated with the VR controller and headset based on their unique radar cross-section (RCS), thereby mitigating the influence of the surrounding environment and enabling accurate VR privacy spying.

\begin{figure}[h]
\centering
    \includegraphics[width=0.6\linewidth]{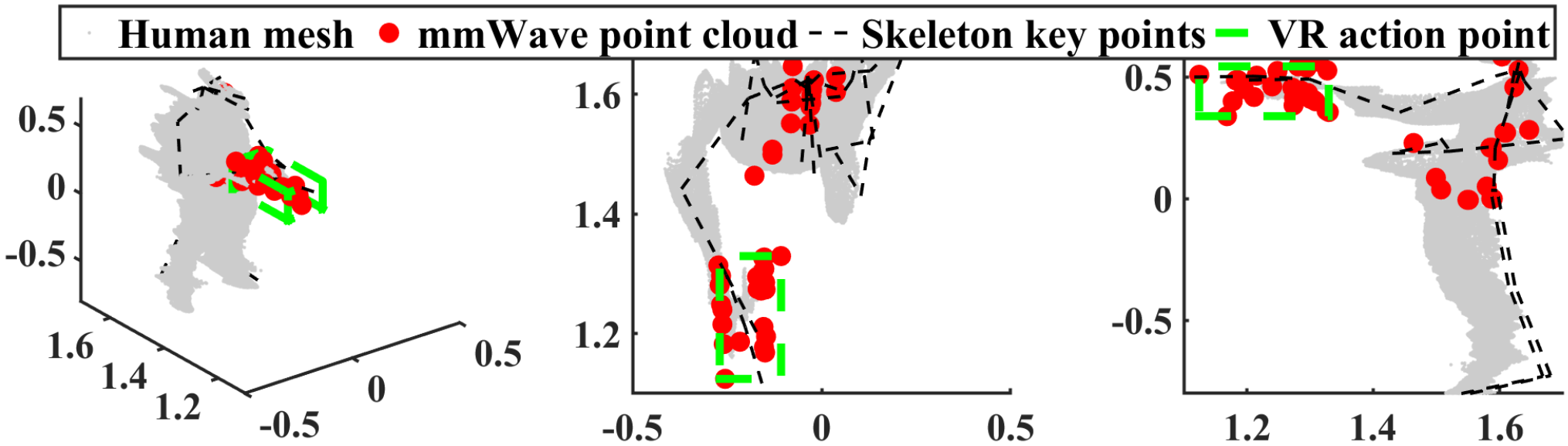}
    \caption{VR action point tracking.}
    \label{fig:motion}
\end{figure}
\vspace{-2mm}

As depicted in Fig.~\ref{fig:motion}, the red points represent the point cloud generated from received mmWave signals, while the green bounding box corresponds to the point cloud with significant VR features. The gray outline represents the human body mesh. The figure indicates that even after penetrating through the obstacles, the mmWave signals retain sufficient information. This information is utilized to recognize VR action and facilitate precise privacy spying. This experiment validates the feasibility of utilizing mmWave signal for VR privacy spying.
\section{System Overview}
The fundamental concept of mmSpyVR is to detect the features of VR actions utilizing mmWave radar, with the ultimate goal of acquiring VR users' private information through VR spying. Focusing on the concealment of this approach, we design mmSpyVR for VR privacy spying without physical and virtual connection with VR users.

\begin{figure}[h]
    \centering
    \includegraphics[width=0.65\linewidth]{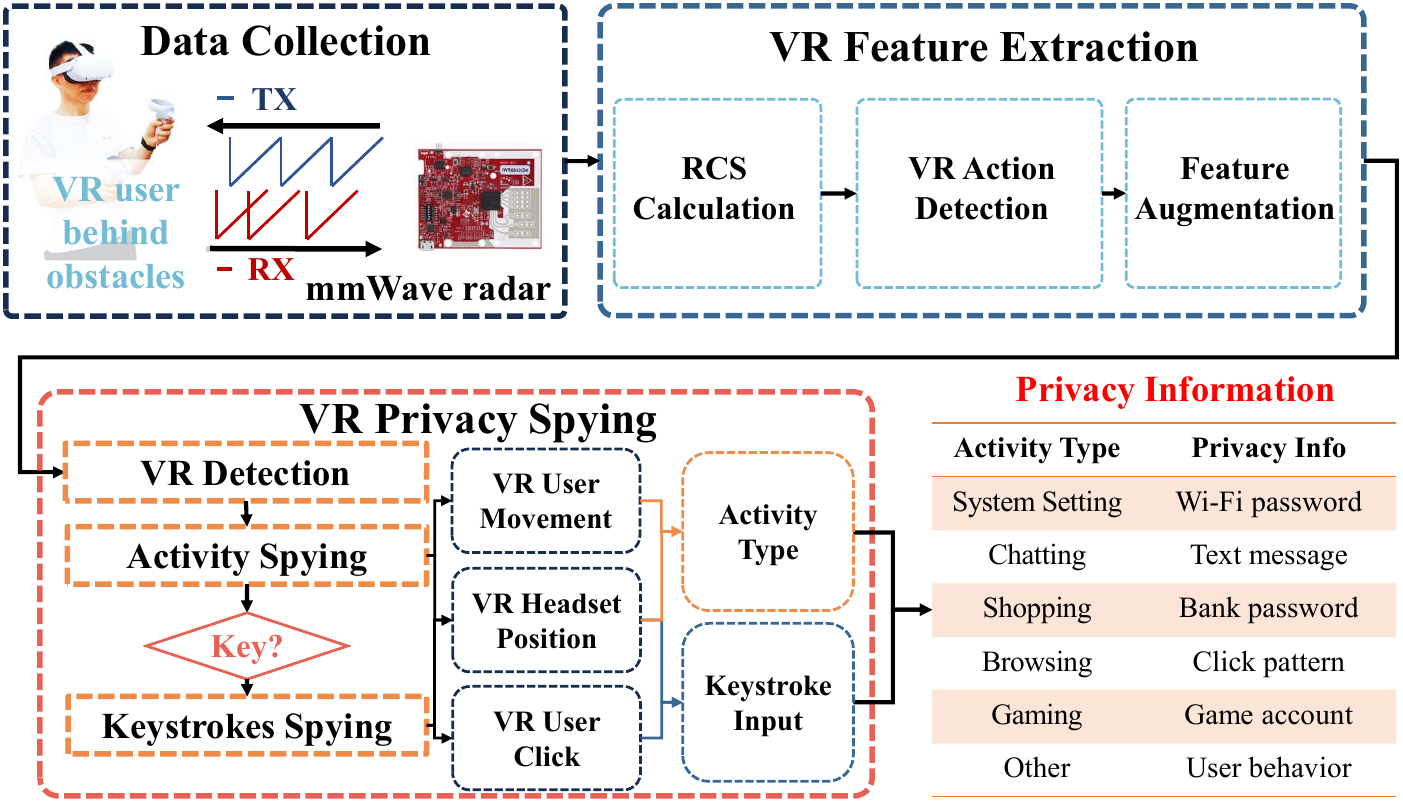}
    \caption{The system overview of mmSpyVR.}
    \label{fig:overview}
\end{figure}

Fig.~\ref{fig:overview} illustrates the design architecture of mmSpyVR, which consists of two main components: VR feature extraction and VR privacy spying. The mmSpyVR gathers the mmWave signals reflected within the area where the VR user is located. These signals, rich with VR action information, are captured by mmWave signal that requires no physical and virtual connection to the VR apparatus. We do not restrict the types of activities users engage in. Participants are encouraged to use VR devices as they would daily, resulting in a diverse and natural range of activities. Through analysis of user behavior patterns, we identify activities leading to privacy leakage, thereby enabling effective privacy spying. The system processes user activities as follows:
\begin{itemize}
    \item In the VR feature extraction module, mmSpyVR utilizes the collected mmWave signals to calculate the radar cross-section (RCS) and create a point cloud embedded with RCS information. After the extraction of RCS, the identified RCS features are utilized for VR action detection to determine whether the user is engaged in ongoing VR activities. A transfer learning-based feature augmentation network is designed to enhance the precision of action spying, leveraging the inherent consistency of VR user action behavior. This network augments the information contained within each frame of VR action data.

    \item The VR privacy spying module leverages augmented point clouds of the user's body for comprehensive activity surveillance. It focuses on activities involving keyboard input, such as settings, shopping, chatting, gaming, and browsing. To handle new activities, a continual learning approach with a model mixture-based method is employed. For keystroke spying, the system uses augmented point cloud data from the VR controller and headset. It incorporates multi-task learning with modules for key press detection and keystroke input prediction. Pre-trained models for various keyboard layouts are utilized, with confidence scores assigned based on spatial distribution, such as click position and the confirmation key press location. For unknown keyboards, a learning process is initiated when confidence falls below a threshold. The system combines recognition of keyboard type, position, and key presses to generate comprehensive keystroke identification outputs, ensuring adaptability across diverse VR environments and input methods.
\end{itemize}

This comprehensive approach for activity analysis and keystroke spying enables the covert acquisition of sensitive privacy information of the VR user, thereby highlighting the vulnerability of privacy intrusion through mmWave radar in VR scenarios while maintaining adaptability to diverse and evolving VR usage patterns.
\section{VR Feature Extraction}
\label{sec4_feasibility}
This section introduces the radar cross-section (RCS), the crucial feature for VR privacy spying. This work first introduces the principle of RCS. Then, we illustrate the methodology of RCS calculation in commodity mmWave radars. After that, we provide a detailed study of VR devices' RCS measurements.

\subsection{Radar Cross-section}
RCS measures the object’s reflection area to radar. A larger RCS indicates that the object is accessible to be detected by radar. The object’s RCS is related to size, material, and orientation. Moreover, RCS features have unique characteristics that bring opportunities for VR privacy spying. (i) First, the RCS of the VR targets is different from its surrounding furniture. (ii) Second, the RCS value is consistent across various sensor-to-target distances. Therefore, utilizing mmWave radars for VR privacy spying introduces new privacy security concerns for VR users, as it reveals sensitive information without their awareness.

\subsection{RCS Calculation on Commodity mmWave Devices} 
RCS measures the echo intensity of a target hit by a radar wave. We utilize the RCS value to infer the target shape. However, off-the-shelf mmWave devices (e.g., TI IWR series~\cite{IWR6843}) do not provide RCS results directly.

The main challenge in RCS calculation is obtaining the signal-to-noise ratio (SNR), the RX average signal and noise strengths ratio. The noise depends on the radar circuit background noise, which is hard to measure on the integrated radar equipment~\cite{10287427}. We assume that RCS is proportional to RX signal strength when other parameters are fixed. We utilize a corner reflector in the calibration phase to get the RX signal strength at different distances $d$ and build the benchmark database $\mathcal{B}(d)$. From $\mathcal{B}(d)$, we get the space Cartesian coordinates $(x, y, z)$, and if the distance $L=\sqrt{x^2+y^2+z^2}$, we compute RX signal intensity $P_{r_t}$ as:

\begin{equation}
    \sigma_p = \frac{P_{r_t}}{\mathcal{B}L}\sigma_r.
\end{equation}

We get $\mathcal{B}(d)$ in the calibration phase. From $\mathcal{B}(d)$, we get the SNR, and then we get the RCS as:

\begin{equation}
    \sigma=\frac{(4\pi)^3 d^4kTF{SNR}}{P_t G_{TX} G_{RX}\lambda^2T_{meas}}.\label{eq_sigma}
\end{equation}

Here $k$ is the Boltzmann constant, $T$ is the antenna temperature, $F$ is the RX noise coefficient, $\lambda$ is the wavelength (constant), $P_t$ is the radar output power, $G_{TX}$ is the TX Antenna Gain, $G_{RX}$ is the RX Antenna Gain, $T_{meas}$ is the measurement time, and $d$ is the target-radar distance, computed from the radar output.

\subsection{RCS Value of VR Targets and Environmental Interference}
Table~\ref{tab:RCS} shows the theoretical RCS values for both VR users and environmental objects, calculated utilizing formulas from book "Radar Cross Section" by Knott~\cite{knott2004radar}. These values reveal a significant difference between VR components and environmental objects. Specifically, environmental fixtures such as the table and television exhibit substantial theoretical RCS values, $60m^2$ and $100m^2$, respectively. In contrast, the VR components that carry significant VR feature information, such as the VR controller and headset, along with the VR user, present theoretical RCS values that are markedly distinct from those of environmental objects, with the VR controller at $5m^2$, the VR headset at $8m^2$, and the VR user at $3.5m^2$. These theoretical values underscore the potential feasibility of identifying VR feature objects from mmWave signal reflections.

\begin{table}[h]
        \caption{Theoretical RCS values of VR target and environmental interference.}
        \centering
        \begin{tabular}{|c|c|c|c|}
            \hline
       \textbf{Type} & \textbf{Group} & \textbf{Target Object} & \textbf{Theoretical RCS}
        \cr
            \hline
        \multirow{3}*{Environmental Interference} & \multicolumn{3}{c|}{\textbf{Environmental Target}} \\
        \cline{2-4}
         ~ & 1 & Table & $60m^2$   \\
        \cline{2-4}
         ~ & 1 & Television & $100m^2$  \\
            \hline
        \multirow{5}*{VR Target} & \multicolumn{3}{c|}{\textbf{VR Devices}}   \\
            \cline{2-4}              
         & 2 & VR controller & $5m^2$               \cr
            \cline{2-4}
         & 2 & VR headset & $8m^2$           \cr
            \cline{2-4}
        ~ & \multicolumn{3}{c|}{\textbf{VR User Body}}   \cr
            \cline{2-4}
         & 3 &  VR user  & 3.5$m^2$      \cr
            \hline
        \end{tabular}
        \label{tab:RCS}
    \vspace{-1mm}
\end{table}

To verify these theoretical predictions and assess the practical feasibility of using RCS for VR privacy spying, we conduct experiments utilizing the commercial mmWave radar IWR6843-ODS. As shown in Fig.~\ref{fig:scenario}, we set up an experimental scenario where VR devices, users, and environmental objects coexist in the same space. We divide our measurements into three groups: the first focuses on measuring the RCS of environmental interference objects, the second measures the RCS of VR devices, and the third measures the RCS of VR users. We utilize the mmWave radar to measure the actual RCS values across different ranges and angles of view for each group, allowing us to compare the RCS characteristics of VR-related objects with those of environmental interference.

The signal from the mmWave radar is sensitive to an object’s posture, indicating a strong correlation between the object’s viewing angle and its RCS value. To quantify this effect, we undertake a series of experiments to assess the effects of viewing angle on RCS. These measurements are conducted in a realistic VR scenario where the user freely utilizes the VR equipment in a room with obstacles.

\begin{figure}[h]
\centering
    \begin{minipage}[c]{0.8\linewidth}
        \centering
        \begin{minipage}[c]{0.56\linewidth}
        \centering
            \includegraphics[width=\linewidth]{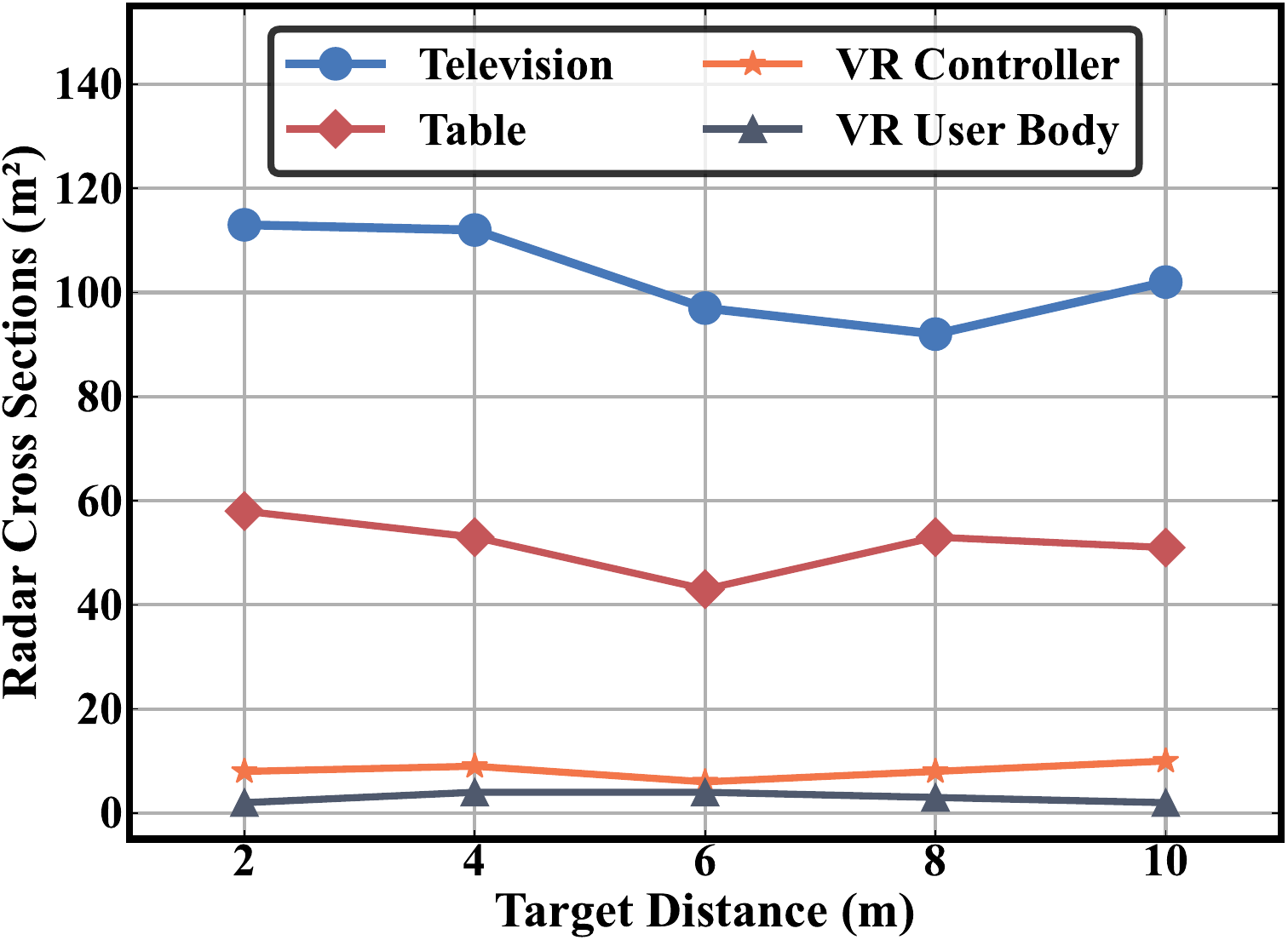}
            \captionof{figure}{RCS at different distances.}
            \label{fig:RCSdistance}
        \end{minipage}
        \hfill
        \begin{minipage}[c]{0.38\linewidth}
        \centering
            \includegraphics[width=\linewidth]{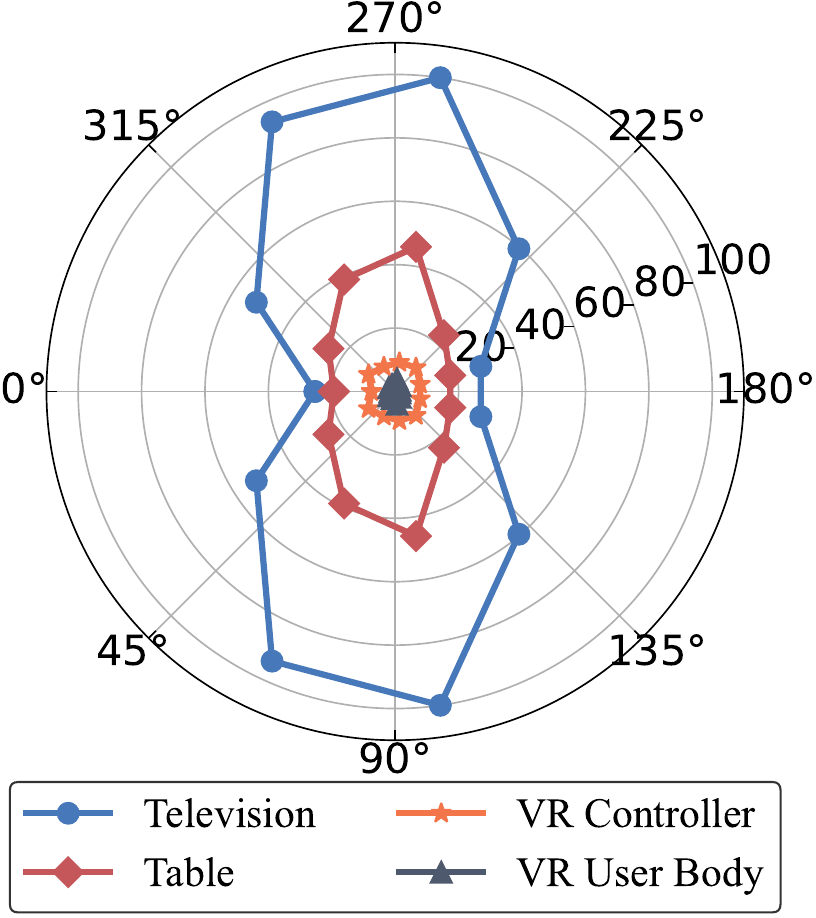}
            \caption{RCS at different angles of view.}
            \label{fig:RCSangle}
        \end{minipage}
    \end{minipage}
\end{figure}

\textbf{Experimental RCS at Different Distance.}
First, to eliminate other influencing factors, we confirm that an object’s RCS is not affected by distance. We demonstrate the RCS of the VR user and the visual interference when the sensor is at different distances, moving from 2 meters to 10 meters, from the object. As shown in Fig.~\ref{fig:RCSdistance}, the RCS of the VR user body and the VR controller is very stable in the range of 2$\sim$4$m^2$ and 6$\sim$10$m^2$, respectively. However, the RCS of the table and television is in the range of 43$\sim$58$m^2$ and 92$\sim$113$m^2$. Therefore, the RCS value of the VR target differs from the environmental objects. Moreover, the RCS value of the same target at different distances is generally consistent, which supports the theory that the target's RCS, in the sensing coverage of mmWave radar, is not influenced by the target distance.

\textbf{Experimental RCS at Different Angles of View.}
\label{subsec_RCS}
Furthermore, as illustrated in Fig.~\ref{fig:RCSangle}, we collect RCS data for the VR user and the environmental objects across different angles. The angles of view refer to the azimuth angles measures by rotating the mmWave radar around the target object in a 360-degree circle. We observe that the objects with VR features exhibit different RCS values at various angles compared with environmental objects. For instance, the RCS values for the VR controller range from $7.50m^2$ to $10.00m^2$, while the RCS of the VR user’s body is between $1.60m^2$ and $4.00m^2$. These contrasts are stark compared to the higher RCS values of environmental objects, which exhibit values up to $100.00m^2$ and $46.08m^2$, respectively. This distinct difference in RCS values across different angles allows us to identify objects with VR features based on their RCS signatures.

\vspace{-2mm}
\vspace{-2mm}
\section{VR Privacy Spying}
\textbf{Preliminaries}
The input is a multi-frame point cloud tensor of size $s \times N \times d$, where $s$ represents the combined batch and sequence length, $N$ is the number of points per frame, and $d$ is the dimensionality of each point, including the six features: $x$, $y$, $z$, velocity, intensity, and RCS. The output is a tensor of size $s \times N \times d_{feature}$, where $d_{feature}$ is the dimension of the feature vector for each point. The feature computation process is as follows:

\begin{equation}
\vec{y}_i = \sum_{\vec{x}_j \in \mathcal{X}} \vec{a}_{ij} \odot \sigma(\vec{x}_j, RCS_j),
\end{equation}

\noindent where $\vec{y}_i$ is the feature vector for the $i$-th point, $\vec{a}_{ij}$ is the feature attention weight between points $i$ and $j$, and $\sigma(\vec{x}_j, RCS_j)$ is a function that calculates the feature score of point $j$ incorporating its RCS value. The feature attention weight $\vec{a}_{ij}$ is derived by applying feature transformations that include the RCS values:

\begin{equation}
\vec{a}_{ij} = \tau \left(\eta \left(\theta \left(\phi(\vec{x}_i, RCS_i), \chi(\vec{x}_j, RCS_j)\right)+\xi \right)\right),
\end{equation}

\noindent where $\phi$ and $\chi$ are feature transformations that integrate the RCS values of points $i$ and $j$, $\theta$ is a relation function, $\eta$ is the mapping function (typically an MLP), $\xi$ is the learned bias, and $\tau$ is the non-linear activation function. This formulation utilizes the RCS feature for the privacy spying process.

\subsection{VR Detection}
This section introduces the VR action detection and feature augmentation module, which utilizes VR devices' distinct Radar Cross Section (RCS) feature to identify VR-related activities and enhance the extracted features based on the detected VR actions.

\subsubsection{VR Action Detection}
\label{subsec_handtracking}
To identify the mmWave point cloud data corresponding to VR actions, we propose an algorithm that exploits the unique RCS characteristics of VR devices. The primary goal of this algorithm is to distinguish and extract the point clouds originating from the VR controller and headset by filtering out the point clouds with RCS values that do not match the expected values for these devices. By identifying the presence of VR devices in the mmWave point cloud data, mmSpyVR determines whether the user is engaged in VR-related activities instead of other non-VR actions.

\begin{algorithm}[h]
    \caption{Extract mmWave Point Cloud on VR Controller and Headset}
    \label{alg.Extract mmWave Point Cloud on VR Controller and Headset}
    \begin{algorithmic}[1]
    \renewcommand{\algorithmicrequire}{\textbf{Input:}}
    \renewcommand{\algorithmicensure}{\textbf{Output:}}
    \REQUIRE $signal$, mmWave Signal $ptcloudData$, Point Cloud
    \ENSURE $ptcloud$, Point Cloud on Controller and Headset
    \FOR{$i = 1$ to $end_frame$}
    \STATE $RCS \leftarrow \text{calculation}(signal)$
    \STATE $ptcloud \leftarrow ptcloudData[i]$
    \STATE $Location \leftarrow \text{readLocation}(ptcloud)$
    \STATE $Intensity \leftarrow \text{readField}(ptcloud, Intensity)$
    \STATE $Velocity \leftarrow \text{readField}(ptcloud, Velocity)$
    \STATE $transfered \leftarrow \text{transformCoordinate}(Location)$
    \STATE $filtered \leftarrow \text{filterPointCloud}(transfered, RCS)$
    \STATE $VR \leftarrow \text{DBSCAN}(filtered)$
    \STATE $ptcloud \leftarrow [VR, Intensity, Velocity]$ 
    \ENDFOR
    \end{algorithmic}
\end{algorithm}

Specifically, the pseudocode presented in Algorithm~\ref{alg.Extract mmWave Point Cloud on VR Controller and Headset} outlines the steps for localizing mmWave point cloud data specifically for the VR feature. This process takes advantage of the distinct RCS characteristics of these devices. We take the mmWave point cloud as input. Each input data frame is processed to obtain the point cloud and calculate the RCS values. The location, intensity, and velocity fields are extracted from the point cloud. After that, the coordinate transformation is applied to align the point cloud with the global coordinate system. A point cloud filter is then used, which relies on the RCS values to distinguish and retain the point clouds corresponding to the VR controller and headset. By focusing on the point clouds that exhibit RCS values expected from the metallic components of the VR devices, the algorithm separates these from the rest of the environment. The density-based spatial clustering of applications with noise (DBSCAN) algorithm \cite{DBSCAN} clusters these filtered points, and the final point cloud data, which includes the VR feature clusters along with their intensity and velocity attributes, is stored. This method of using RCS as a distinguishing feature enables precise identification of point clouds emanating from the VR controller and headset.

Subsequently, in the mmSpyVR feature identification phase, we scrutinize the point cloud data generated by the VR user's body to identify areas rich in VR features. The model then aggregates the attention scores of points within close spatial proximity, reflected in the group score equation:

\begin{equation}
G_k = \sum_{\vec{a}_{ij} \in G_k} \vec{a}_{ij} \cdot \eta(RCS_i, RCS_j),
\end{equation}

\noindent where $G_k$ represents the $k$-th group of points, and $\eta(RCS_i, RCS_j)$ is a function that incorporates the RCS values of points $i$ and $j$ into the group score computation. Each group's score represents the VR feature attention.

To determine the most informative points for privacy spying, we introduce a neighborhood global feature identification (NGFI) mechanism, which selects the groups with the highest global attention scores. The global attention score for each group is computed as follows:

\begin{equation}
g_j = \frac{1}{|G_j|} \sum_{\vec{x}_i \in G_j} \vec{y}_i^T \vec{w} \cdot \zeta(RCS_i),
\end{equation}

\noindent where $G_j$ is the $j$-th group of points, $\vec{y}_i$ is the vector attention score for point $\vec{x}_i$, $\vec{w}$ is a learnable weight vector, and $\zeta(RCS_i)$ is a function that incorporates the RCS value of point $i$ into the global attention score computation.

The NGFI mechanism identifies the feature-discriminative region with the highest global attention score group:

\begin{equation}
R_S = \underset{j}{\mathrm{argmax}} \, g_j,
\end{equation}

\noindent where \( R_S \) stands for identified VR feature region. This equation selects the group \( g_j \) that maximizes the global attention score, which is then deemed as the region with the highest relevance to the target feature. This approach ensures that RCS values are fully integrated into the computation of feature scores, thereby determining the importance of each point in the point cloud for VR privacy spying.

\subsubsection{Feature Augmentation}
\label{subsec_augmentation}
Upon detecting that a user is engaged in VR activities, we enhance the VR-specific features. Specifically, the point clouds with VR features are often sparse, challenging the subsequent VR privacy spying. By augmenting the identified point cloud feature, we infuse the sparse frame with the spatial richness of historical data. This augmentation network is uniquely self-supervised, utilizing the inherent consistency of VR user action behavior to guide the augmentation process. By leveraging the predictable nature of these movements, the network teaches itself to refine and enrich the point cloud, ensuring that the augmented output remains true to realistic VR user motion. This self-supervised method not only circumvents the need for labeled datasets but also aligns the augmented data with the intricate patterns of genuine VR user motion, which is critical for the subsequent analysis by ActNet and KeyNet.

\begin{figure}[h]
\centering
    \includegraphics[width=0.7\linewidth]{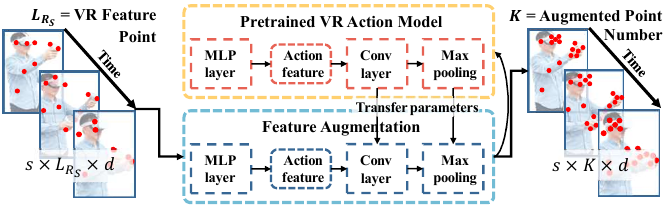}
    \caption{Feature augmentation.}
    \label{fig:augmentation}
\end{figure}

Specifically, the model depicted in Fig.\ref{fig:augmentation} demonstrates that when the density level $L_{R_S}$ of the current VR point cloud frame falls below the threshold $K$, indicating a sparse mmWave point cloud, the pre-trained VR action model is utilized to enhance the frame's features. This model stores known mmWave radar features corresponding to specific VR user actions. The input to this module is the point cloud with identified VR features from the previous stage. It processes the sparse current frame alongside prior denser frames to estimate the VR user's action. Subsequently, the pre-trained action feature model is applied to refine the estimation by augmenting the sparse frame's features with the stored action features. These augmented features are then utilized to predict VR feature points with a network consisting of convolutional layers, max pooling, and linear layers. The output is a point cloud with enhanced density compared to the original sparse input. Furthermore, the feature augmentation network's parameters are updated in real-time via transfer learning to adapt to the user's actions.

\subsection{Activity Spying via Body Action}
\label{subsec_ActNet}
\begin{figure}[h]
    \centering
    \includegraphics[width=0.8\linewidth]{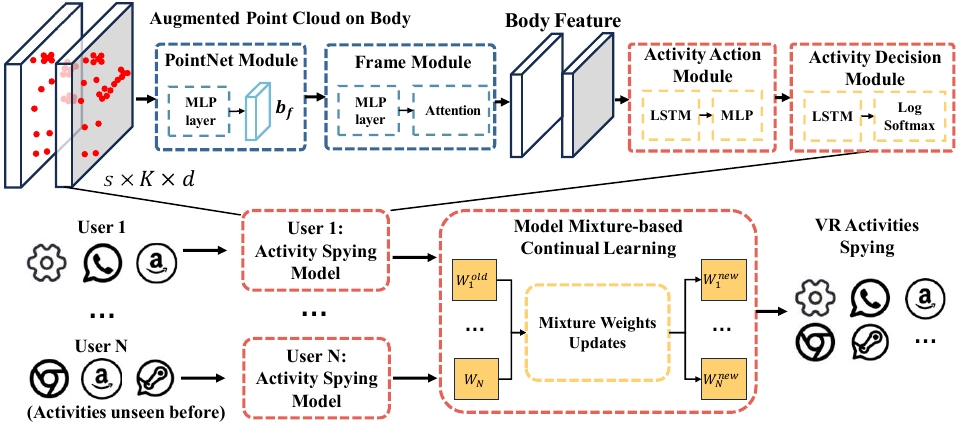}
    \caption{Activity spying network.}
    \label{fig:actnet}
\end{figure}

In this subsection, we introduce an activity spying network, as shown in Fig.~\ref{fig:actnet}, named AppNet, that identifies specific VR activities utilizing augmented feature frames. This attention-based model utilizes point clouds gathered during VR privacy spying to distinguish between different applications based on user actions. We employ a continual learning approach to enhance the model's robustness, allowing it to adapt to new and unknown activities. This method ensures the system's effectiveness in dynamic VR environments by continuously learning from both existing and new data.

Specifically, we employ a continual learning approach with a mixed model method. Our system comprises three main modules: the PointNet and frame module, the action module, and the decision module. The PointNet~\cite{PointNet} and frame module extract distinctive features of the VR user's body from the mmWave point cloud frame, taking point cloud data of size $s \times N \times d$ as input and outputting feature vectors of size $s \times K \times d$. Its formal representation is $f = \sigma(W_f P + b_f)$, $F = \psi(f)$, where $P$ is the augmented point cloud data, $f$ is the body point cloud feature vector, $F$ is the body feature, $\sigma$ is a non-linear activation function, $W_f$ and $b_f$ are learnable weights and biases, and $\psi$ represents the PointNet function. The action module processes the sequential information of the body features, producing motion features that reflect the activity patterns of the user's body. It is represented as $m_t = \tanh(W_m [F_t, h_{t-1}] + b_m)$, where $m_t$ is the motion feature at time step $t$, $F_t$ is the encoded body feature, $h_{t-1}$ is the previous hidden state, and $W_m$ and $b_m$ are learnable parameters. Finally, the decision module classifies the VR user's activities based on these motion features, formulated as $act = \text{softmax}(W_d m + b_d)$, where $act$ is the activity label, and $W_d$ and $b_d$ are learnable weights and biases.

To address previously unseen activities and maintain model robustness, we implement a model mixture-based continual learning approach, as shown in the bottom of Fig.~\ref{fig:actnet}. This method allows the system to adapt to new activities while preserving performance on known ones~\cite{Deep_reinforcement}. The mixture model is represented as $P(act|m) = \sum_{i=1}^{K} \pi_i P_i(act|m)$, where $P(act|m)$ is the probability of an activity given motion feature $m$, $K$ is the number of mixture components, $\pi_i$ are mixture weights, and $P_i(act|m)$ are individual model probabilities. When an unseen activity is encountered, the system updates the mixture weights and individual models: $\pi_i^{new} = (1 - \alpha)\pi_i^{old} + \alpha P_i(act|m)$ and $W_d^{new} = W_d^{old} + \eta \nabla_{W_d} \log P_i(act|m)$, where $\alpha$ and $\eta$ are the learning rate for mixture weights and model parameters, and $W_d$ are the decision module weights. This continual learning approach enables the system to adapt to new activities based on newly collected data. The updated decision module then incorporates the new activity: $act = \text{softmax}(W_d^{new} m + b_d)$. This adaptive mechanism ensures that the model classifies known and unseen activities, maintaining performance across a diverse range of VR interactions.

\subsection{Keystroke Spying via Controller and Headset Action}
\begin{figure}[h]
    \centering
    \includegraphics[width=1\linewidth]{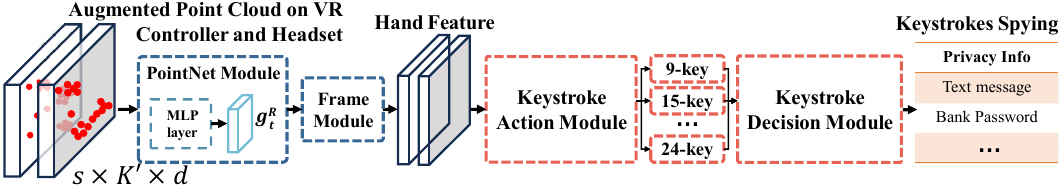}
    \caption{Keystroke spying network.}
    \label{fig:keynet}
\end{figure}

The VR keystroke spying module processes augmented point clouds with dimensions $s \times K' \times d$, where $K'$ represents the number of point clouds on the VR controller and headset in each augmented frame. As illustrated in Fig.~\ref{fig:keynet}, the keystroke spying network comprises four main components: the PointNet module, the frame module, the keystroke action module, and the keystroke decision module. This network employs a multi-task learning strategy to identify critical presses, detect the keyboard's position and type, and perform keystroke spying.

Initially, the PointNet module extracts features from the mmWave point cloud data, which are then refined by the frame module using an attention mechanism: $\vec{f} = g_t^R(\psi(\mathbf{P}))$, where $\mathbf{P}$ is the augmented point cloud input, $\psi$ is the PointNet function, and $\vec{f}$ is the enriched feature vector. The keystroke action module then analyzes these features to detect key press actions using a Bi-LSTM: $\vec{m}t = \tanh(W_m [\vec{f}t, \vec{h}{t-1}, \vec{h}{t+1}] + b_m)$, where $\vec{m}t$ represents the motion feature output at time step $t$.

Upon detecting a key press, the keystroke decision module employs pre-trained models for various keyboard layouts (e.g., 9-key, 15-key, 24-key) to predict the user's input. Each prediction is assigned a confidence score based on both the spatial distribution of the user's key press actions and the confirmation key press location. The confidence score for each keyboard layout $i$ is calculated as: $C_i = \text{softmax}(W_c \cdot D_i + W_f \cdot F_i + b_c)$, where $D_i$ is the spatial distribution of key presses mapped to layout $i$, $F_i$ is the feature vector representing the confirmation key press location for layout $i$, and $W_c$, $W_f$, and $b_c$ are learnable parameters. The final keystroke prediction is determined by: $\mathbf{key} = \arg\max_i(C_i \cdot \text{softmax}(W_d \vec{m} + b_d))$, where $\mathbf{key}$ is the identified keystroke output, and $W_d$ and $b_d$ are learnable weights and biases from the decision module. This approach allows the system to adapt to different keyboard configurations and provide accurate predictions based on the most likely layout.

In cases where the system encounters an unknown keyboard distribution, resulting in low confidence scores across the known layouts, the keystroke decision module initiates a learning process for potential new layouts. Specifically, when the confidence scores fall below a predefined threshold $\tau_{confidence} = 0.5$~\cite{Privacy_Leakage1, Privacy_Leakage2, Privacy_Leakage3}, the system records the spatial distribution of key presses and initiates a new layout learning procedure. This adaptive mechanism is incorporated into the multi-task learning objective: $\mathcal{L}_{total} = \mathcal{L}_{key} + \lambda (\mathcal{L}_{pos} + \mathcal{L}_{type} + \mathcal{L}_{conf} + \mathcal{L}_{new})$, where $\mathcal{L}_{new}$ represents the loss associated with learning new layouts, and $\lambda$ is a weighting factor balancing the contributions of each task. This approach ensures that the system remains flexible and adapt to novel keyboard configurations encountered in diverse VR environments.
\section{Evaluation}
\label{sec6_evaluation}

\subsection{Experimental Setup}
\label{sec6_implementation}
In this section, we discuss the hardware implementation and experimental configuration of mmSpyVR, emphasizing its performance in various real-world scenarios and alignment with the design principles.

\begin{figure}[h]
\centering
    \begin{subfigure}[t]{0.28\linewidth}
        \includegraphics[width=\linewidth]{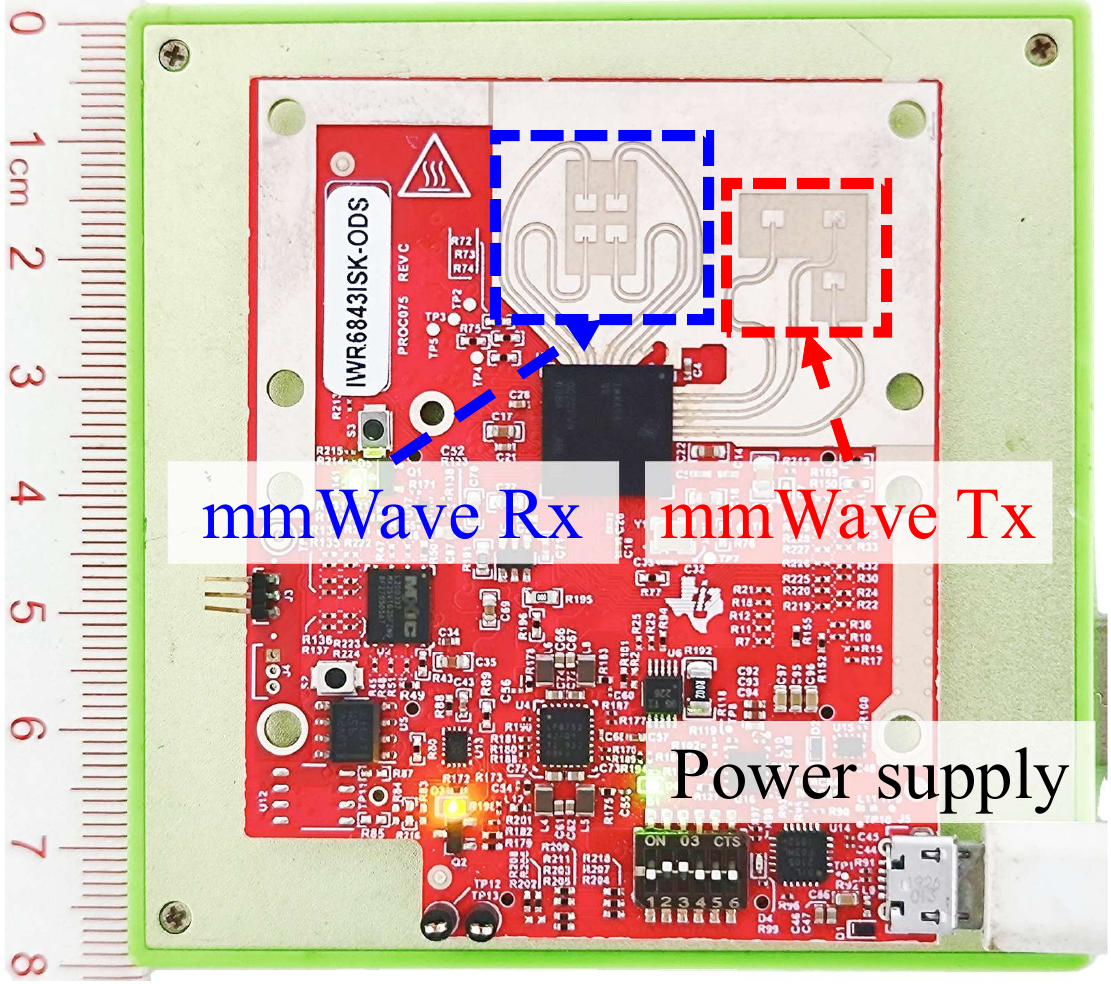}   
        \caption{VR privacy spying device}
        \label{fig:attackdevice}
    \end{subfigure}
    \hspace{5mm}
    \begin{subfigure}[t]{0.42\linewidth}
        \includegraphics[width=\linewidth]{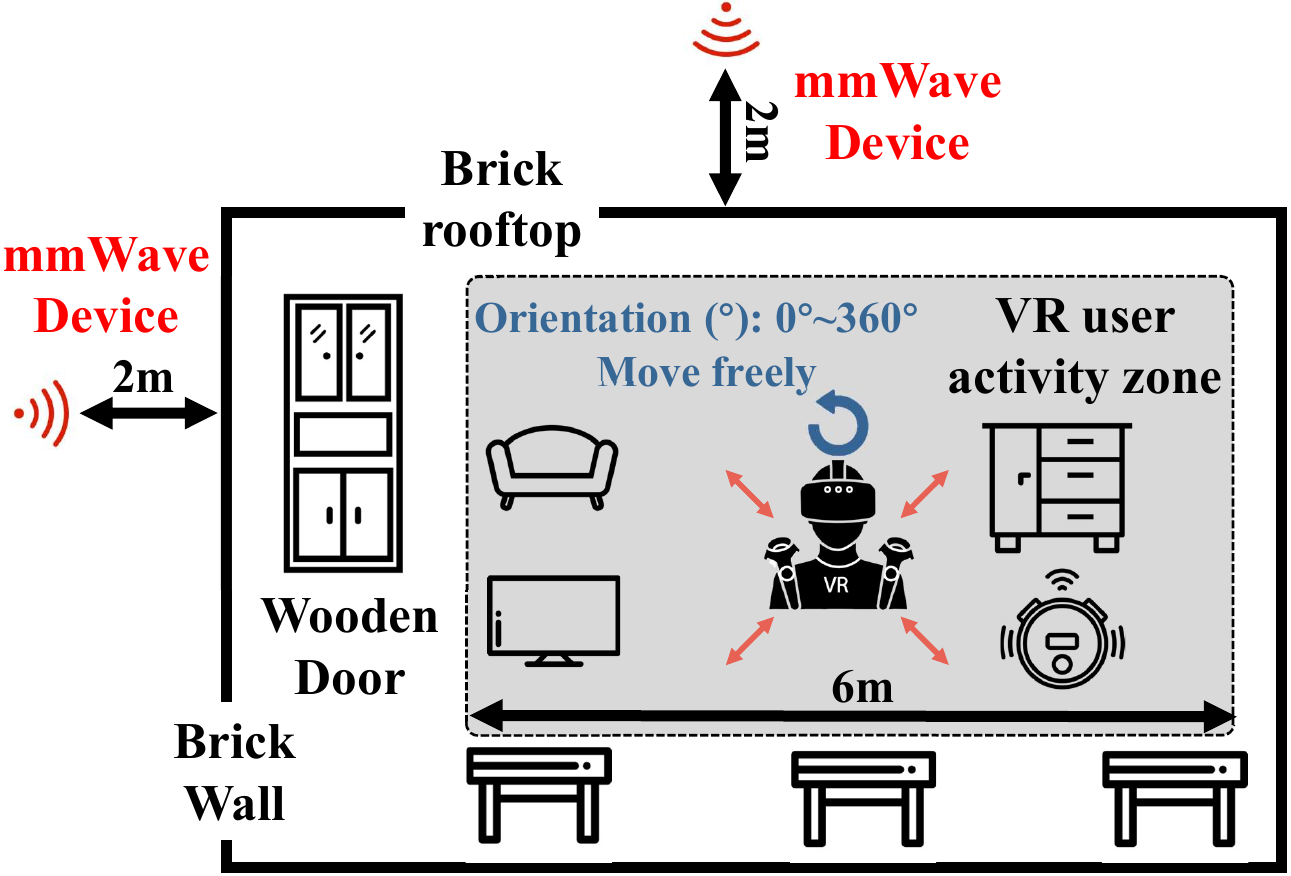}
        \caption{\textbf{Experiment topology.}}
        \label{fig:exptopolog}
    \end{subfigure}
\caption{The implementation of mmSpyVR.}
\label{fig:implementation}
\end{figure}

\label{subsec_attackdevice}
\textbf{VR privacy spying device.} 
As shown in Fig.~\ref{fig:attackdevice}, the system device is designed to focus on low power consumption and compactness. Its primary function is gathering mmWave data from VR users, with the data processing performed offline. The VR privacy spying device is concealed within a power bank, with an overall weight of 96 grams and dimensions of 8 cm in length and width. The mmWave radar device utilized in our experiments is the IWR6843-ODS. This device is configured with a bandwidth of 4000.14 MHz and operates at a frequency of 60 GHz. It utilizes a 3TX and 4RX antenna array, which provides a distance resolution of about 0.75 cm and a velocity resolution of about 0.24 m/s. The radar is set up in a self-transmitting and receiving configuration with an RX Gain of 30 dB. This covert integration enables the placement of the spying device in various settings, such as home and office centers, without attracting attention. This design is primarily for data collection in mmSpyVR, as the nature of the VR spying task requires capturing user behavior and privacy data without being detected, making real-time data processing unnecessary and impractical.

\textbf{Topology.}
Fig.~\ref{fig:exptopolog} illustrates our experimental setup, designed to replicate real-world scenarios. The VR user moves within a 10-meter room with typical household items while the mmWave device is positioned outside the room and on the rooftop. This configuration allows the spying device to penetrate walls and doors, capturing data as the user moves. The presence of furniture introduces realistic obstructions, e.g., sofa, cabinet, and television, further challenging the system. We collect mmWave data across distances ranging from 2 to 8 meters at various angles, ensuring a comprehensive evaluation of mmSpyVR's capabilities in authentic environments.

We do not restrict the types of user behaviors and focus on activities involving keyboard input, such as settings, shopping, chatting, gaming, and browsing. A continual learning approach with a model mixture-based method is employed to handle new activities. Additionally, mmSpyVR is adaptable to various types of activities and keyboard layouts. By leveraging augmented point clouds and heatmaps of VR actions, mmSpyVR recognizes different keyboard layouts and performs accurate keystroke spying. This ensures comprehensive surveillance and adaptability across diverse VR environments and input methods.

\textbf{Data Collection.}
Our study involves 22 participants, 11 males and 11 females, aged between 21 and 58. We assess the effectiveness of the proposed privacy spying through user studies approved by our institutional review board (IRB 2021ZDSYLL089-P01). We collected a total of 3,600 data sets, total of 4,320,000 frames, amounting to 12 TB of data, with 100 sets in each category. Each set includes both mmWave IQ and point cloud data. To ensure robust and generalizable results, we randomly split the collected data into three distinct subsets: 70\% of the data is used for training, 10\% for validation, and 20\% for testing.

\textbf{System Setup.}
Our experiments encompass a range of popular VR systems, including Meta Quest 2, Sony PlayStation VR, and HTC VIVE XR. These diverse platforms enabled us to collect a broad spectrum of data across different VR hardware configurations. Following data collection, mmSpyVR conducts its VR spying analysis on the aggregated dataset. The spying system operates on an Asus ROG laptop running Windows 11, equipped with a GeForce RTX 4060 GPU, providing the necessary computational power for our privacy spying algorithms.

\textbf{Implementation.}
Our mmSpyVR system consists of three key components: VR action detection, activity spying, and keystroke spying. The VR action detection module uses 16×16 patches to partition point clouds, with input sequences of 25 frames optimized for keystroke recognition. Activity spying employs PointNet and LSTM networks, while keystroke spying uses a modified architecture with an embedding network, a 2-layer bidirectional LSTM, and a fully connected layer. We train the model for 700 epochs with early stopping with a patience of 200 epochs to prevent overfitting. Parameter sharing across embedding, LSTM, and FC layers ensures consistency. Our implementation adheres to the original design, enabling accurate analysis of VR interactions across various scenarios and devices.

\textbf{Evaluation Matrix.}
In our comprehensive evaluation, we compare mmSpyVR with state-of-the-art (SOTA) methods in both point cloud modeling and mmWave radar sensing domains. The SOTA methods include point cloud models such as Point Transformer~\cite{PointTransformer} and Point 4D Transformer~\cite{Point4DTransformer}, as well as mmWave radar models RadarNet~\cite{RadarNet} and Tesla rapture~\cite{Tesla-Rapture}. We evaluate the performance of mmSpyVR and the SOTA methods in various scenarios, comparing their accuracy and performance in spying VR privacy.

Specifically, our evaluation of mmSpyVR's performance encompasses multiple perspectives. In Section~\ref{subsec_overallperform}, we present the overall performance of mmSpyVR on three VR devices under four obstacle scenarios. Section~\ref{subsec_sota} compares mmSpyVR with the existing SOTA methods and evaluates their performance in VR privacy spying. In Section~\ref{subsec_key}, we measure the system's accuracy in recognizing keystrokes for the 26 letters and 10 digits. In Section~\ref{subsec_app}, we examine the precision with which the system identifies the type of activity the VR user utilizes, including gaming, chatting, shopping, browsing, and system settings. In Section~\ref{subsec_scenarios}, we test the robustness of the system under various challenging scenarios, such as penetrating a wooden door, the orientation from the side (0°, 90°, 180°, and 270°) to the top, and the distance from 2 to 8 meters. In Section~\ref{subsec_information}, we evaluate the system's effectiveness in recovering VR user's privacy. To evaluate the performance of mmSpyVR, we conducted experiments to demonstrate its capability to recognize three types of VR devices under four different obstacle conditions: Unobstructed, Wooden Door, Brick Wall, and both Wooden Door and Brick Wall.

\subsection{Overall Performance}
\label{subsec_overallperform}
We evaluate the performance of mmSpyVR in terms of activity type spying accuracy and keystroke spying accuracy across four obstacle scenarios. 

\begin{figure}[h]
\centering     
    \captionof{table}{Experiment devices, obstacles, and results.}
    \label{tab:result}
    \resizebox{0.7\linewidth}{!}{
        \begin{tabular}{lccc}
        \toprule
        \textbf{Obstacles} & \textbf{VR Device} & \textbf{Activity Type Spying} & \textbf{Keystrokes Spying} \\
        \midrule
        \multirow{3}{*}{Combined} & Meta Quest 2 & 80.3\% & 75.8\% \\ 
         & Sony PlayStation VR & 78.2\% & 73.6\% \\ 
         & HTC VIVE XR & 82.9\% & 76.2\% \\  
        \midrule
        \multirow{3}{*}{Brick Wall} & Meta Quest 2 & 86.6\% & 82.9\% \\ 
         & Sony PlayStation VR & 83.2\% & 80.1\% \\ 
         & HTC VIVE XR & 85.8\% & 81.5\% \\ 
        \midrule
        \multirow{3}{*}{Wooden Door} & Meta Quest 2 & 90.2\% & 86.1\% \\ 
         & Sony PlayStation VR & 88.3\% & 84.6\% \\ 
         & HTC VIVE XR & 90.1\% & 90.2\% \\ 
        \midrule
        \multirow{3}{*}{Unobstructed} & Meta Quest 2 & 98.5\% & 91.8\% \\ 
         & Sony PlayStation VR & 97.3\% & 91.3\% \\ 
         & HTC VIVE XR & 98.2\% & 93.5\% \\ 
        \bottomrule
        \end{tabular}
    }
\end{figure}

Table~\ref{tab:result} summarizes the overall performance of mmSpyVR for each VR device under different obstacle conditions. The activity spying accuracy refers to the \textit{top-1} accuracy of classifying activity types, and the keystroke spying accuracy represents the \textit{top-1} accuracy of identifying the 26 alphabetical and 10 numerical keys on the virtual keyboard. The system accurately identifies the activity type and the keystroke input. These reveal the user's private and sensitive data. These experimental results highlight the capability of mmSpyVR to perform privacy spying on VR users even under obstructed conditions. In the following sections, we present the performance of mmSpyVR in various scenarios, covering its ability to spy activity types, keystrokes, and private information.

\subsection{Comparison with SOTA}
\label{subsec_sota}
\begin{figure}[h]
    \centering
    \begin{minipage}[c]{0.48\textwidth}
        \includegraphics[width=\linewidth]{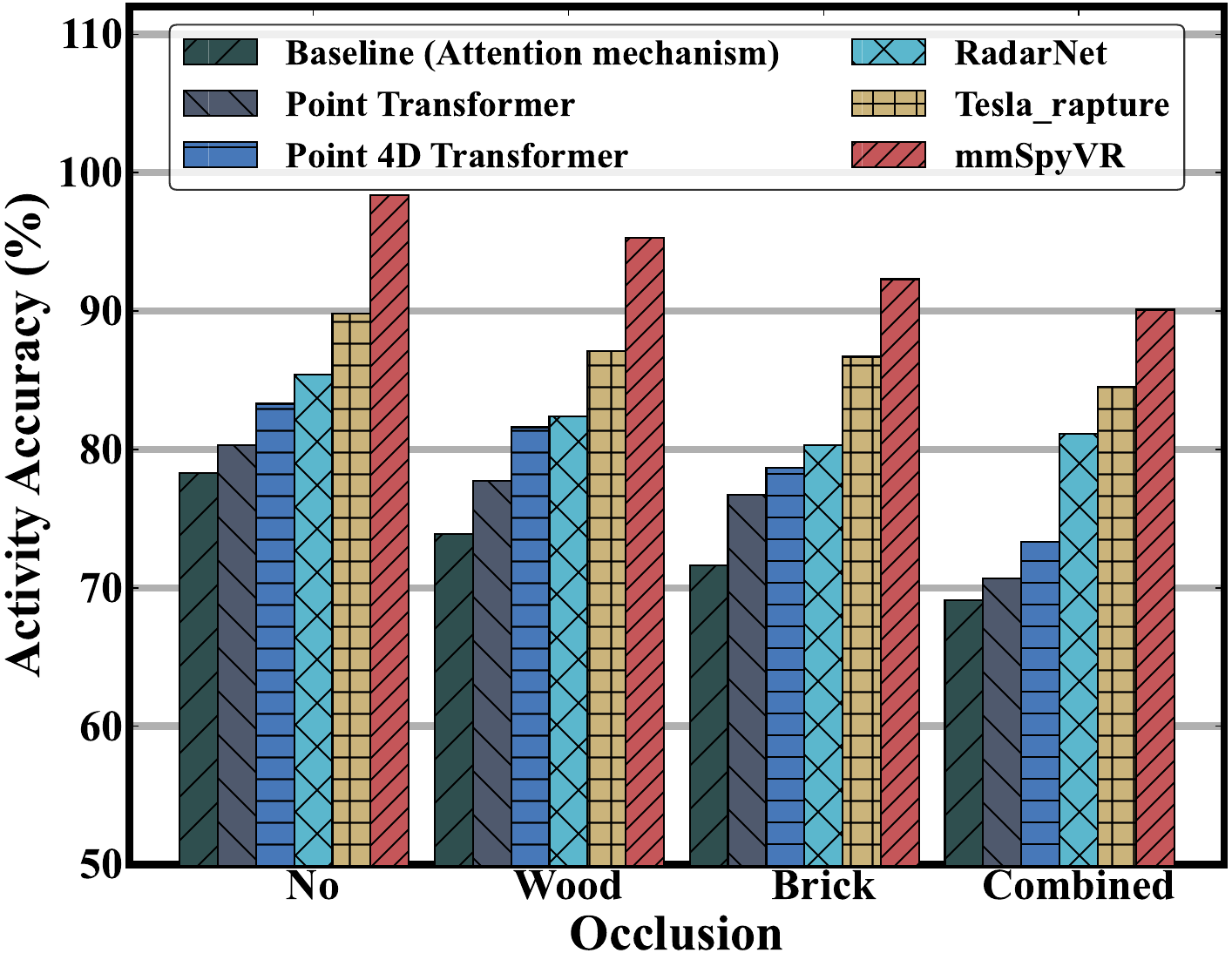}
        \caption{Activity spying vs. SOTA.}
        \label{fig:activity}
    \end{minipage}
    \hfill
    \begin{minipage}[c]{0.48\textwidth}
        \includegraphics[width=\linewidth]{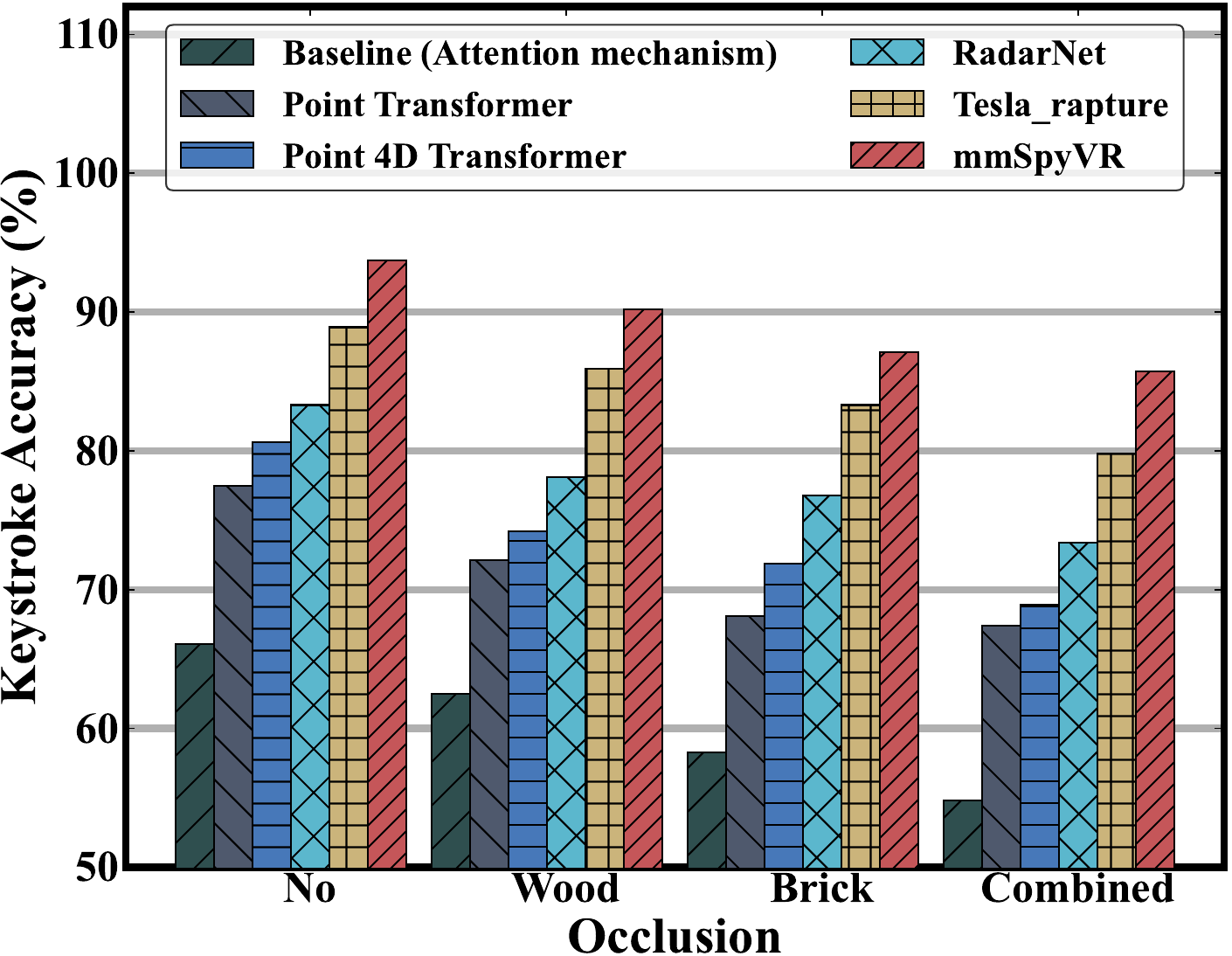}
        \caption{Keystroke spying vs. SOTA.}
        \label{fig:keystroke}
    \end{minipage}
\end{figure}

Fig.~\ref{fig:activity} and~\ref{fig:keystroke} illustrate the average performance of mmSpyVR compared to other state-of-the-art (SOTA) methods for activity and keystroke spying, respectively, across different VR devices under various occlusion scenarios. The SOTA methods include point cloud models such as Point Transformer~\cite{PointTransformer} and Point 4D Transformer~\cite{Point4DTransformer}, as well as mmWave radar gesture spying models RadarNet~\cite{RadarNet} and Tesla rapture~\cite{Tesla-Rapture}. Regarding the comparison with mmSpyVR and SOTA models, we ensured a fair comparison by utilizing the same raw IQ signal data as input. It's worth noting that while RadarNet~\cite{RadarNet} utilizes range-Doppler data as input, and our mmSpyVR utilizes point cloud data, we are able to generate both types of data from the original IQ signals. As described in our "Principles of mmWave Sensing" section, we processed the raw IQ signals to produce range-Doppler data for RadarNet and point cloud data for our mmSpyVR model. This approach allowed us to train and compare both models utilizing data from the same source, ensuring a fair and accurate comparison of their respective accuracies.

Specifically, for activity spying, mmSpyVR achieves an average top-1 accuracy of 98.4\% with no occlusion and maintains a high average accuracy of 90.1\% even in combined occlusion scenarios across different VR devices. In comparison, for the average performance across different VR devices, the best-performing SOTA method, Tesla rapture, only reaches an accuracy of 89.8\% without occlusion and drops to 84.5\% under combined occlusion. The Point Transformer and Point 4D Transformer achieve accuracies of 80.3\% and 83.3\% without occlusion, respectively, and their performance declines under occlusion conditions.

For keystroke spying, mmSpyVR demonstrates superior average performance across VR devices compared to SOTA methods. Without occlusion, mmSpyVR achieves a remarkable 93.7\% accuracy, while Tesla rapture, the best-performing SOTA method, reaches 88.9\%. Under combined occlusion, mmSpyVR maintains an impressive accuracy of 85.7\%, whereas the performance of other methods degrades. RadarNet, for instance, achieves an accuracy of 83.3\% without occlusion but drops to 73.4\% under combined occlusion. The Point Transformer and Point 4D Transformer exhibit accuracies of 77.5\% and 80.6\% without occlusion, respectively, and their performance also decreases notably under occlusion scenarios. These results highlight the robustness and effectiveness of mmSpyVR in handling various occlusion conditions for both activity and keystroke spying tasks, outperforming SOTA methods from point cloud modeling and mmWave radar gesture recognition domains.

\subsection{Illustration of Various Keyboard}
\begin{figure}[h]
    \centering
    \includegraphics[width=0.65\linewidth]{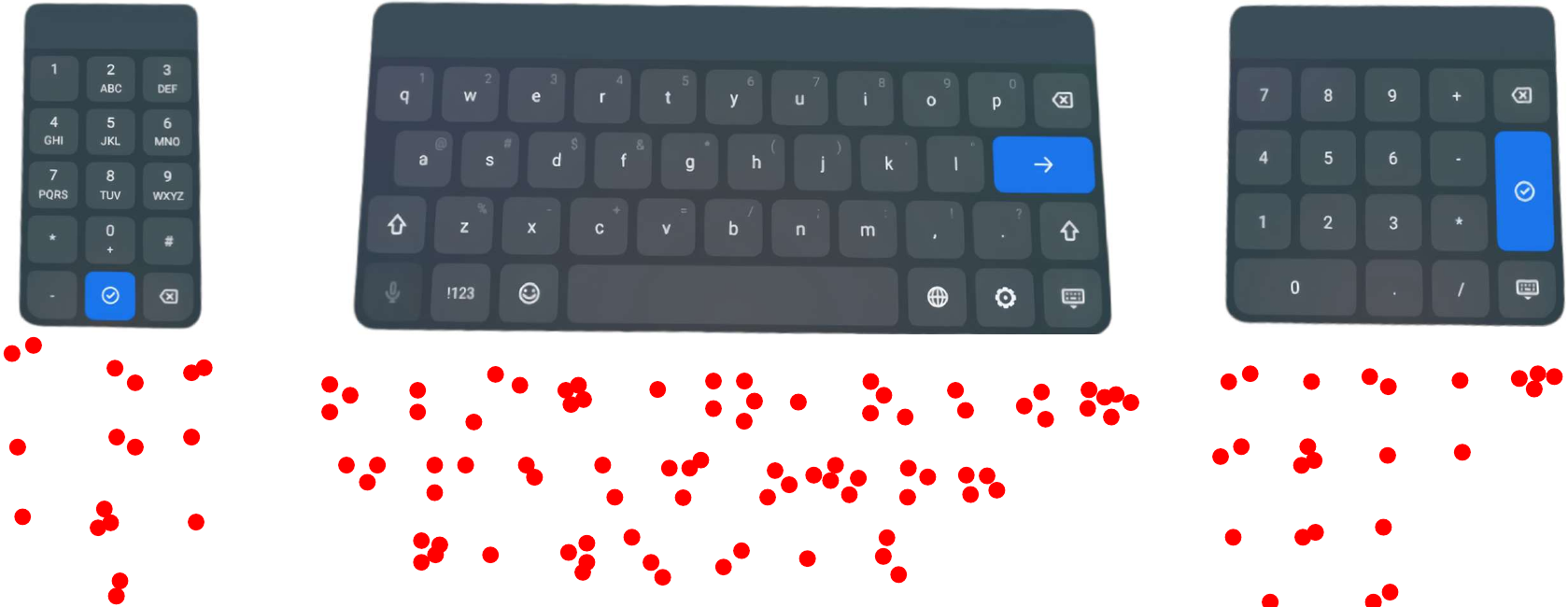}
    \caption{Keystroke click distribution.}
    \label{fig:click}
\end{figure}

The distribution of user clicks in VR applications provides valuable insights into the type of keyboard layout being utilized. As shown in Figure \ref{fig:click}, the 2D projection of click positions over the course of a user's interaction with a VR application clearly reflects the underlying keyboard layout. Since changing layouts mid-use disrupts the user experience and requires re-entering information, VR users do not change keyboard layouts during a single session. This stability allows us to identify and analyze the keyboard layout in use throughout the session.

Fig.~\ref{fig:click} illustrates three distinct keyboard layouts commonly used in VR applications: a 24-key layout, a 9-key layout, and a 15-key layout. The click distribution for the 24-key layout shows a clear rectangular pattern with multiple rows, typical of a full QWERTY keyboard. The click distribution for the 24-key layout resembles a typical computer keyboard with multiple rows forming a clear rectangular pattern. The 9-key layout shows a compact 3x3 grid with an additional “0” key below. The 15-key layout features a 3x3 grid on the right side and mathematical operators on the bottom and right sides. These distinct click patterns help us accurately identify the keyboard layout in use, which is essential for precise keystroke spying and analysis in VR environments.

\subsection{Performance of Keystroke Spying}
\label{subsec_key}
\begin{figure}[h]
    \centering
    \includegraphics[width=\linewidth]{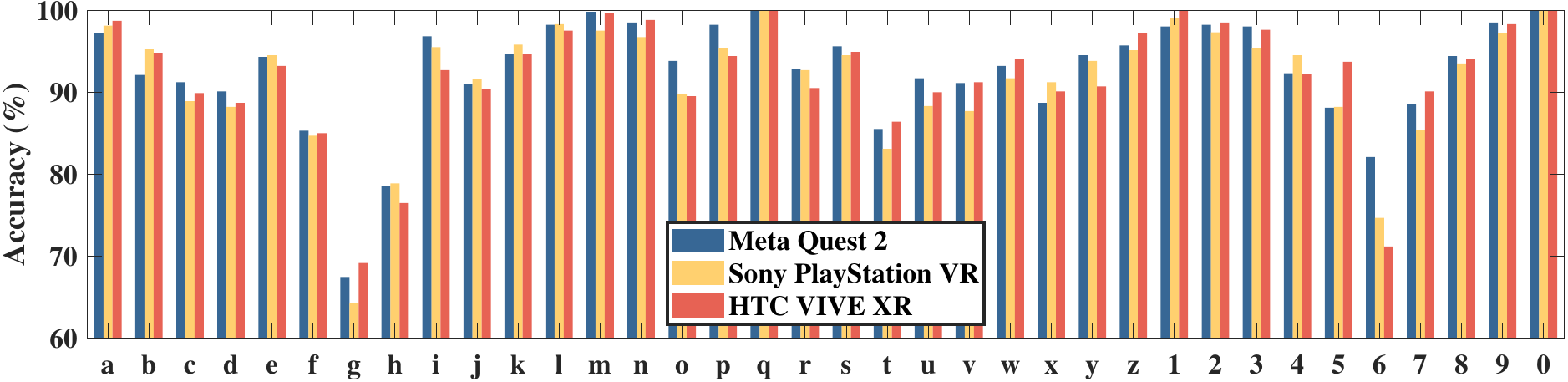}
    \caption{Keystrokes spying accuracy.}
    \label{fig:keybar}
\end{figure}

To demonstrate the performance of mmSpyVR, we subsequently conducted experiments utilizing the most challenging 24-key keyboard. Experimental results illustrated in Fig.~\ref{fig:keybar} reveal the precision with which mmSpyVR discerns individual keystrokes of the VR device. The histogram indicates a notable correlation between the accuracy of keystroke spying and the key's placement on the virtual keyboard. 

Specifically, edge keys such as 'm', 'q', 'z', '1', and '0' exhibit exceptionally high spying accuracies, averaging 99\%, 100\%, 100\%, 99\%, and 100\%, respectively. The distinct spatial separation of these keys from the cluster of other keys contributes to their prominent characteristics in the coordinate dimensions, making them less susceptible to confusion with neighboring keys and yielding higher spying accuracy. In contrast, keys situated centrally on the keyboard like 'g', 't', and '6' face a higher chance of being incorrectly identified as adjacent keys. Their average spying accuracies are lower, standing at 67\%, 85\%, and 76\%, respectively. Despite these challenges, mmSpyVR demonstrates a robust ability to identify VR users' interactions with virtual keyboard keys. The subsequent confusion matrix depicts mmSpyVR's performance on keystroke spying.

\begin{figure}[h]
\centering
    \includegraphics[width=0.55\linewidth]{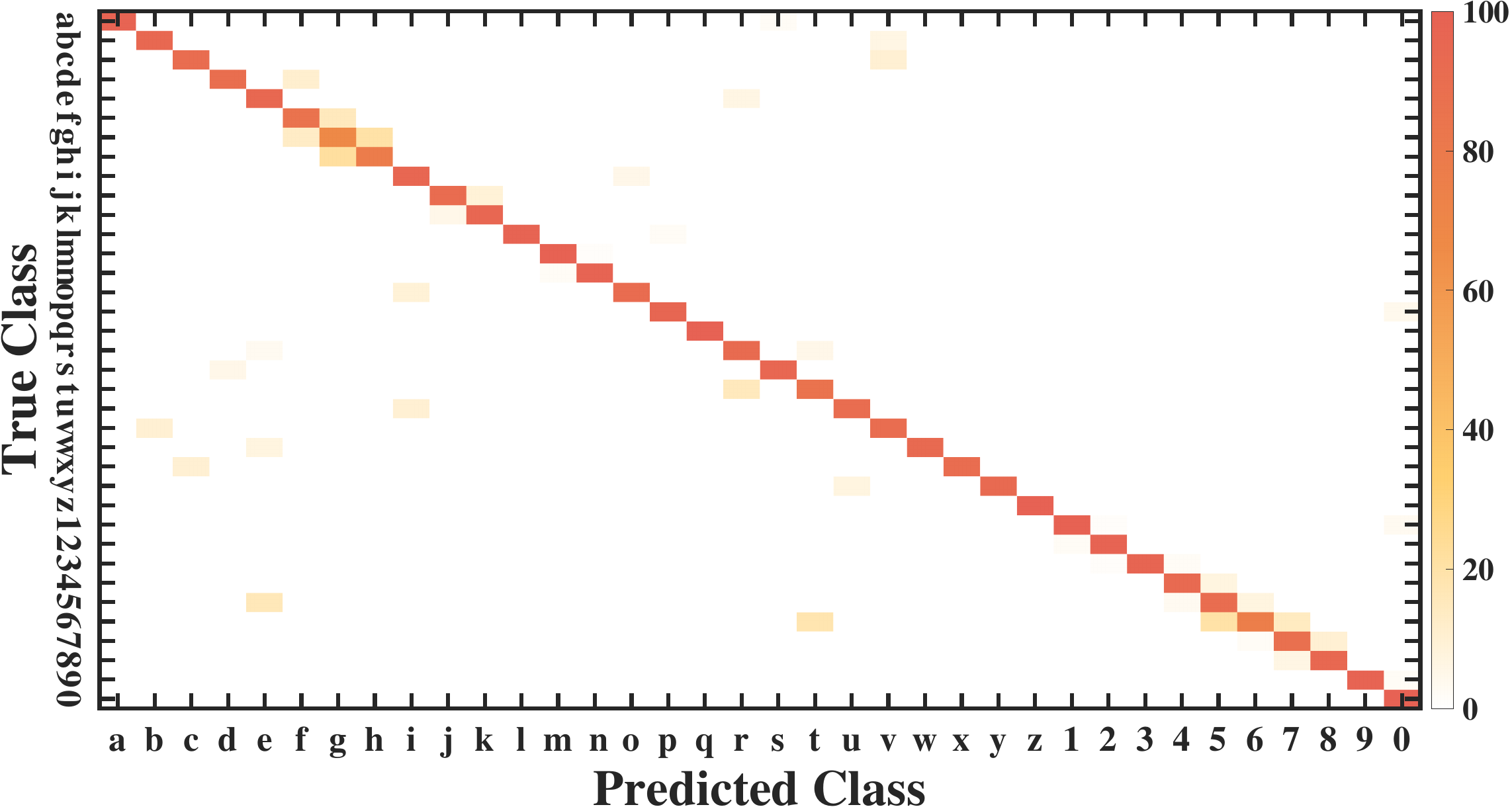}
    \captionof{figure}{Keystrokes spying confusion matrix.}
    \label{fig:keymatrix}
\end{figure}

Fig.~\ref{fig:keymatrix} presents the confusion matrix for keystroke spying, offering an insightful depiction of mmSpyVR's capability in identifying VR user keystrokes. This matrix utilizes identical axes for both "True Class" and "Predicted Class", representing the 26 alphabetical and 10 numerical keys. The minimal and evenly scattered off-diagonal instances indicate a low probability of misclassification. Keys on the virtual keyboard's periphery, such as 'a', 'z', '1', and '0', show higher accuracy due to their distinct positions. Central keys like 'f', 'g', and 'h' exhibit slightly lower accuracy due to potential confusion with adjacent keys. Overall, this confusion matrix corroborates mmSpyVR's proficiency in accurate keystroke spying across the virtual keyboard layout.

\subsection{Performance of Activity Spying}
\label{subsec_app}
We employ mmSpyVR to perform  on three of the mainstream VR devices: Meta Quest 2, Sony PlayStation VR, and HTC VIVE XR aiming to recognize the activity types that the VR users engage with. We focused on five types of VR activities, i.e., Gaming, Chatting, Shopping, Browsing, and System Setting.

\begin{figure}[h]
    \centering
    \begin{minipage}[c]{0.48\textwidth}
        \includegraphics[width=\linewidth]{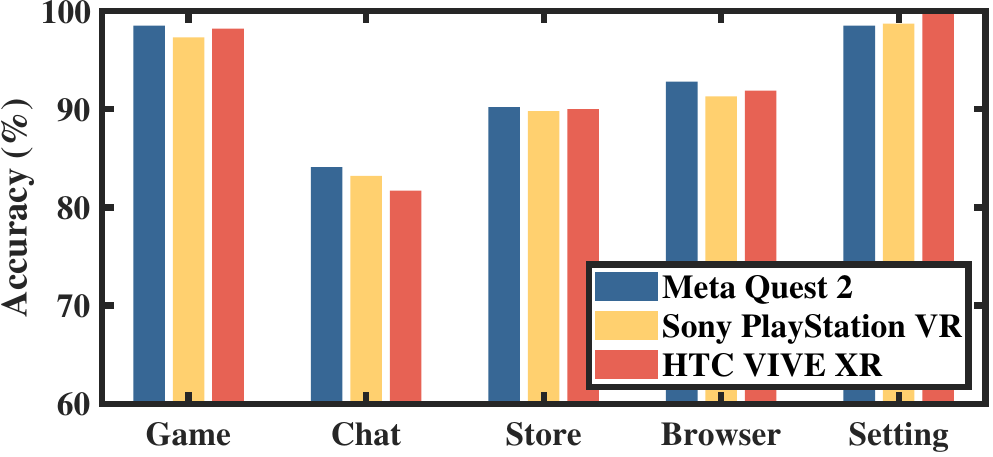}
        \caption{Activity spying accuracy.}
        \label{fig:appbar}
    \end{minipage}
    \hfill
    \begin{minipage}[c]{0.48\textwidth}
        \includegraphics[width=\linewidth]{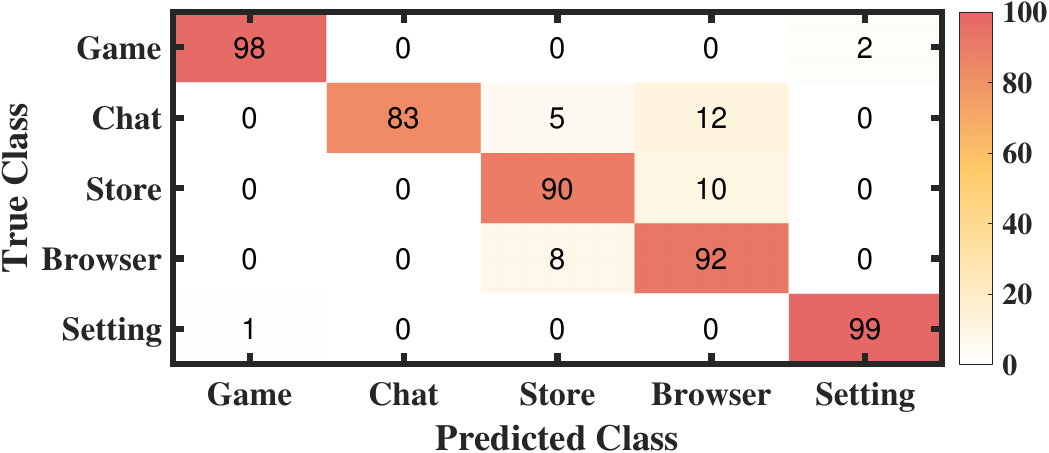}
        \caption{Activity spying confusion matrix.}
        \label{fig:appmatrix}
    \end{minipage}
\end{figure}

The experimental results in Fig.~\ref{fig:appbar} show that mmSpyVR accurately classifies activity types across these devices. The 'Game' and 'System Setting' activities achieved the highest accuracy rates, averaging 98\% and 99\%, respectively. This high level of accuracy is due to the unique user behaviors observed during gaming and system setting adjustments, such as significant body twists and hand swings in gaming and hand clicks in system settings. These features are captured by mmWave radars and learned by mmSpyVR. The average accuracy rates for 'Chat', 'Store', and 'Browser' are 83\%, 90\%, and 92\%, respectively. The similar operational features between 'Store' and 'Browser' led to their lower spying accuracy than 'Game' and 'System Setting'. Additionally, the lower accuracy in 'Chat' is due to less pronounced motion characteristics since VR users move their fingers during chats. The following confusion matrix details mmSpyVR's performance in activity type spying. 

Fig.~\ref{fig:appmatrix} illustrates the confusion matrix of the mmSpyVR’s classification performance across the different VR activity types. The matrix reveals that 'Game' and 'System Setting' activities display markedly distinctive VR features, as evidenced by the minimal misclassifications. Out of 100 instances, 'Game' is misclassified as 'System Setting' only once, and conversely, 'System Setting' is confused with 'Game' just twice. This indicates a strong correlation between the activity type and the unique motion signatures that mmSpyVR detects and classifies with high accuracy. For activities with similar motion patterns, such as 'Store' and 'Browser', the confusion between the two is pronounced. Specifically, within 100 instances, 'Store' is mistaken for 'Browser' on 10 occasions, while 'Browser' is incorrectly identified as 'Store' 8 times, reflecting the challenge in distinguishing between the two due to their operational resemblances. In the case of 'Chat' activities, where user movement is subtle and limited to finger motion on the VR controllers' joystick, the matrix shows a higher rate of misclassification, with 12 instances being wrongly classified as 'Browser' and 5 as 'Store'.

\subsection{Performance of Various Scenarios}
\label{subsec_scenarios}
In this section, we conduct a detailed analysis of mmSpyVR's capabilities under various scenarios. Our research dissects three principal factors critical to the success of assaults on VR users: the orientation of the VR user, the type of obstacles, and the distance between the attack device and the victim. We investigate the system's precision when facing four distinct obstacle scenarios, including unobstructed, wooden doors, brick walls, and combinations of wooden doors and brick walls. Additionally, the robustness of mmSpyVR is evaluated under five different orientations of the VR user, i.e., side view with 0$^\circ$, 90$^\circ$, 180$^\circ$, and 270$^\circ$, and top view. Furthermore, we investigate the system's sensitivity to the distance between the attack device and the victim, ranging from 2 to 8 meters. The experimental results demonstrate that mmSpyVR has high resilience in the above scenarios.

\subsubsection{Impact of obstacles}
\begin{figure}[h]
    \centering
    \begin{minipage}[c]{0.33\textwidth}
        \includegraphics[width=\linewidth]{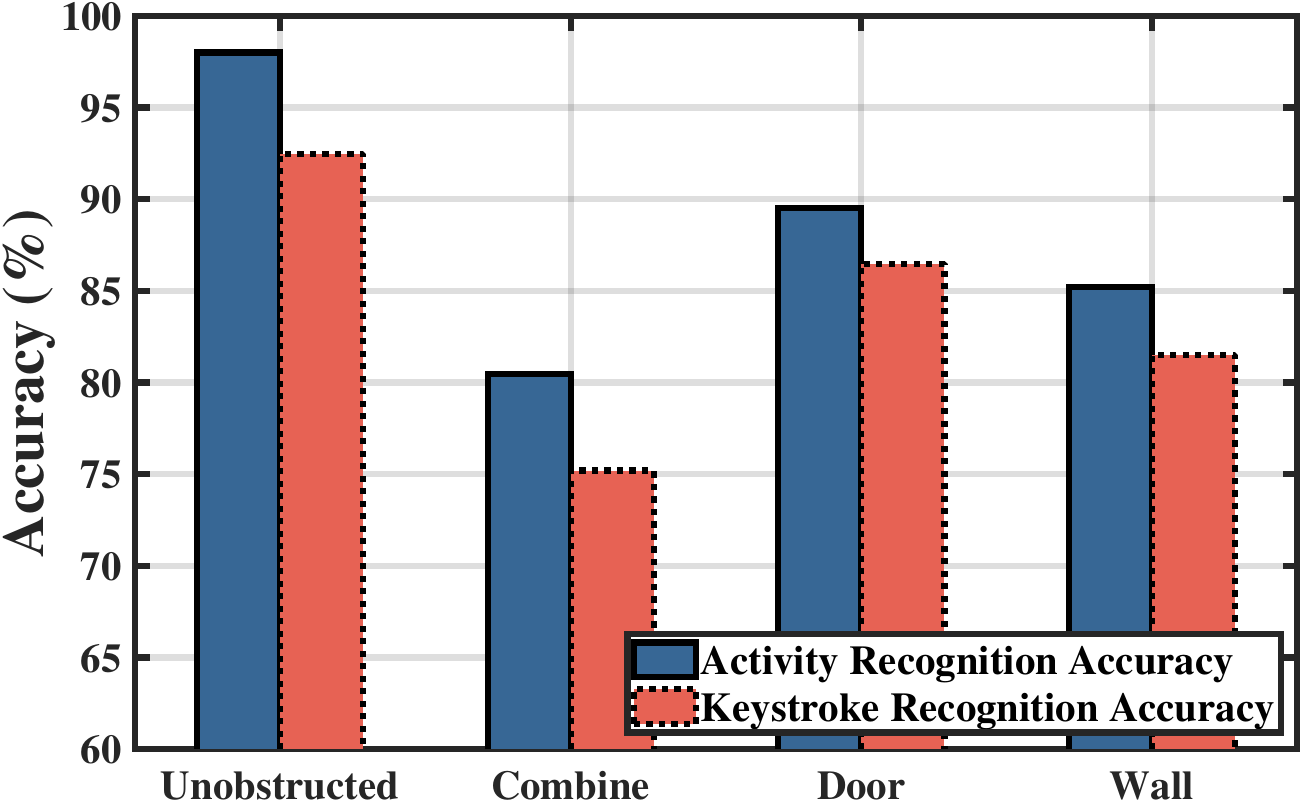}
        \caption{Impact of obstacles.}
        \label{fig:obstacles}
    \end{minipage}
    \hfill
    \begin{minipage}[c]{0.33\textwidth}
        \includegraphics[width=\linewidth]{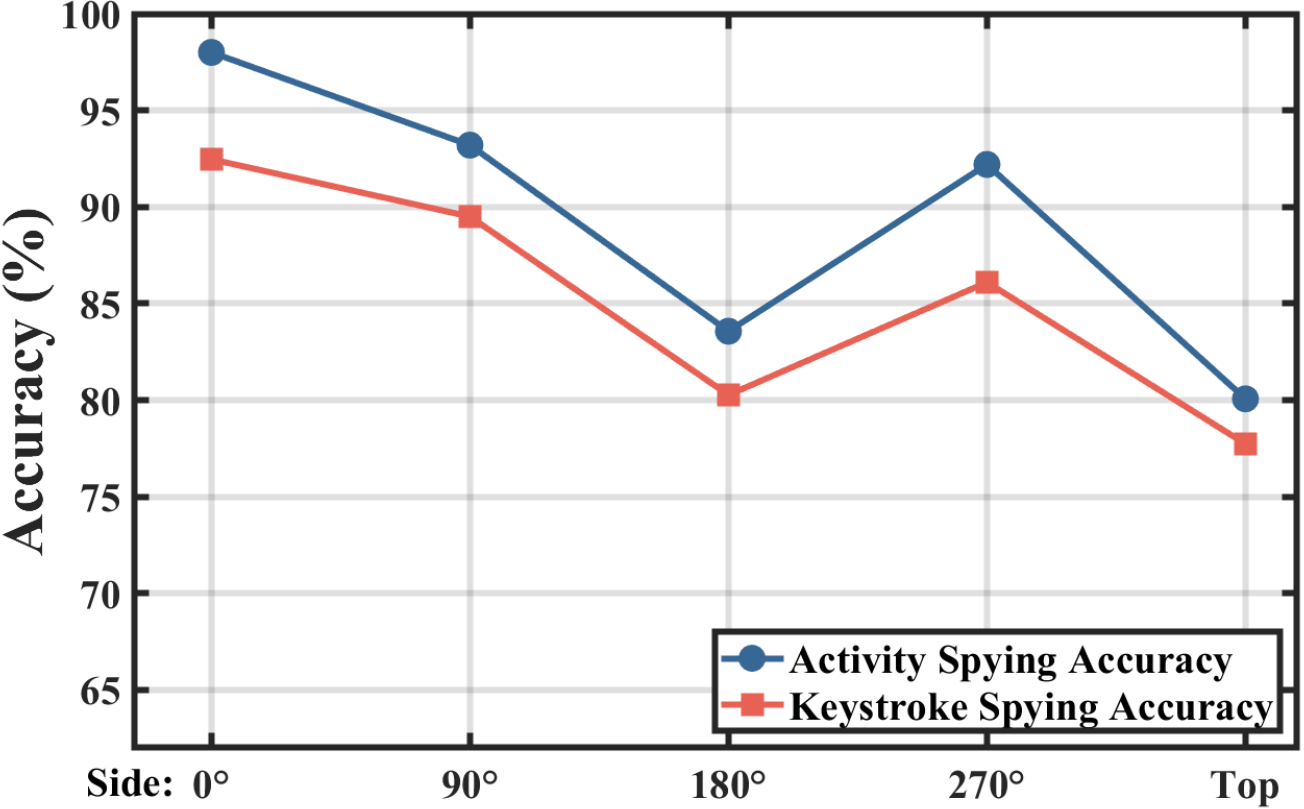}
        \caption{Impact of orientation.}
        \label{fig:orientation}
    \end{minipage}
    \hfill
    \begin{minipage}[c]{0.33\textwidth}
        \includegraphics[width=\linewidth]{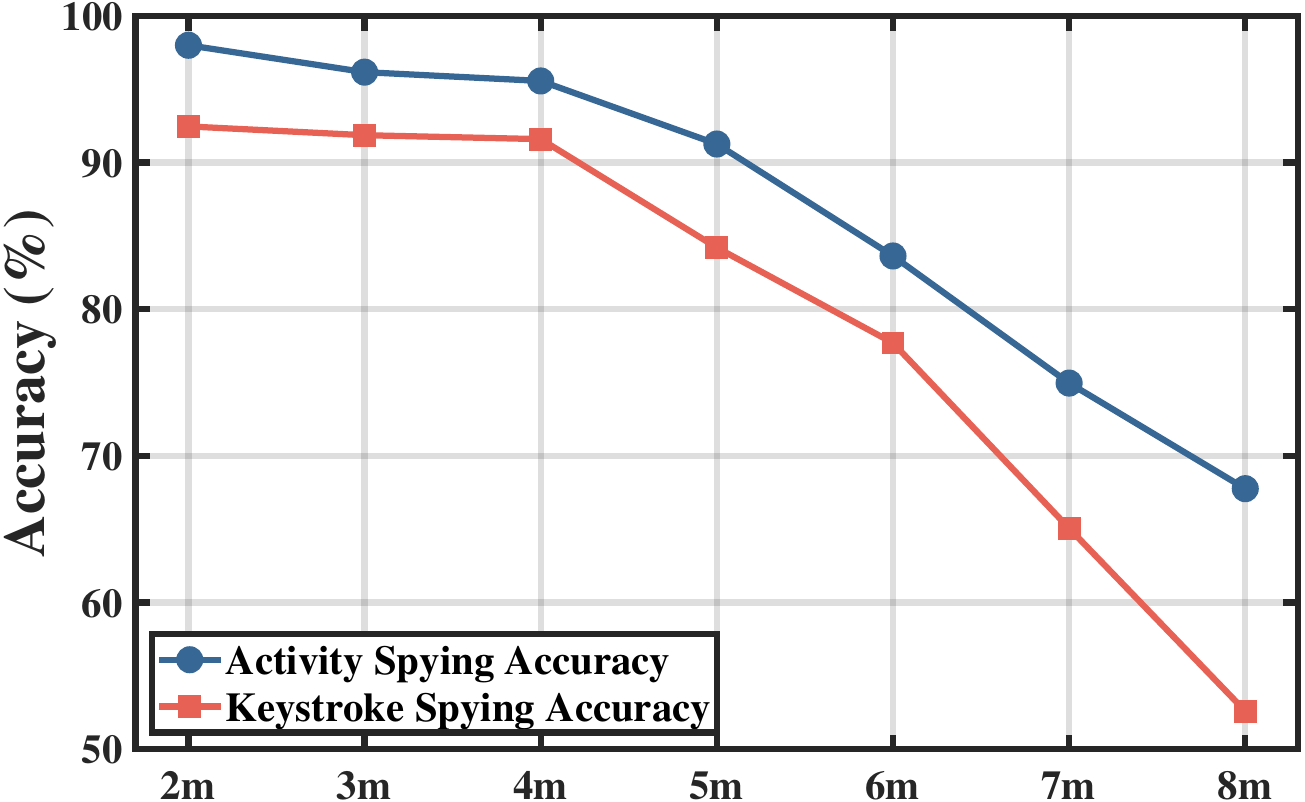}
        \caption{Impact of distance.}
        \label{fig:distance}
    \end{minipage}
\end{figure}

Fig.~\ref{fig:obstacles} illustrates the performance of mmSpyVR under various obstacle scenarios. In the unobstructed scenario, the average activity spying accuracy reached a remarkable 98.5\%. This performance experiences a decline across the different obstruction scenarios, recording averages of 89.5\% with a wooden door, 85.2\% with a brick wall, and 80.5\% when both wood and brick are encountered. The keystroke spying accuracy followed a similar trend, with averages of 92.5\% for the unobstructed, 86.5\% for the wooden door, 81.5\% for the brick wall, and 76.7\% for the combination of both. These results demonstrate the robustness of mmSpyVR.

\subsubsection{Impact of orientation}
Fig.~\ref{fig:orientation} provides a detailed examination of mmSpyVR's accuracy in activity and keystroke spying as VR users adopt different viewing angles. The system's robustness in various orientation scenarios is presented, with side views encompassing 0$^\circ$, 90$^\circ$, 180$^\circ$, and 270$^\circ$, and the top view. In side view orientations, mmSpyVR achieved high average accuracies in activity spying: 98.0\% at 0$^\circ$, 92.9\% at 90$^\circ$, 83.5\% at 180$^\circ$, and 92.2\% at 270$^\circ$. Keystroke spying accuracy also exhibited robust performance with averages of 92.5\% at 0$^\circ$, 89.5\% at 90$^\circ$, 80.3\% at 180$^\circ$, and 86.1\% at 270$^\circ$. Notably, the reduction in accuracy at 180$^\circ$ is attributed to the VR user's back facing the attack device, resulting in the reflection of mmWave signals by the user's back. The top view orientation, which offered a distinct perspective, displayed an activity spying accuracy of 80.1\% and a keystroke spying accuracy of 77.7\%. This perspective presented challenges due to the smaller radar cross-section of the user's hands compared to side views. 

\subsubsection{Impact of distance}
Fig.~\ref{fig:distance} presents an analytical depiction of how distance affects the mmSpyVR's privacy spying accuracy. The trend indicated by the experimental result is quite pronounced. Activity spying accuracy is at its peak at 2m, reaching 98\%. This accuracy then shows a consistent decrement, dropping to 95.6\% at 4 meters, 83.6\% at 6 meters, and falling further to 67.7\% at an 8-meter distance. Keystroke spying accuracy exhibits a correspondingly declining trend, starting strong at 92.5\% for 2m and then descending to 91.6\%, 77.7\%, and finally to 52.6\% as the distance extends. Such performance decline is ascribed to the exponential decrease in the strength of the mmWave signal as the distance elongates. The mmWave device employed in the study, specifically the IWR6843, is engineered to function within an operational range of 8 meters~\cite{IWR6843}. Our mmSpyVR demonstrates its capacity to execute effective VR privacy spying on VR users within this operational threshold.

\subsection{Performance of Private Information Recovery}
\label{subsec_information}
\begin{figure}[h]
    \centering
    \begin{minipage}[c]{0.48\textwidth}
        \includegraphics[width=\linewidth]{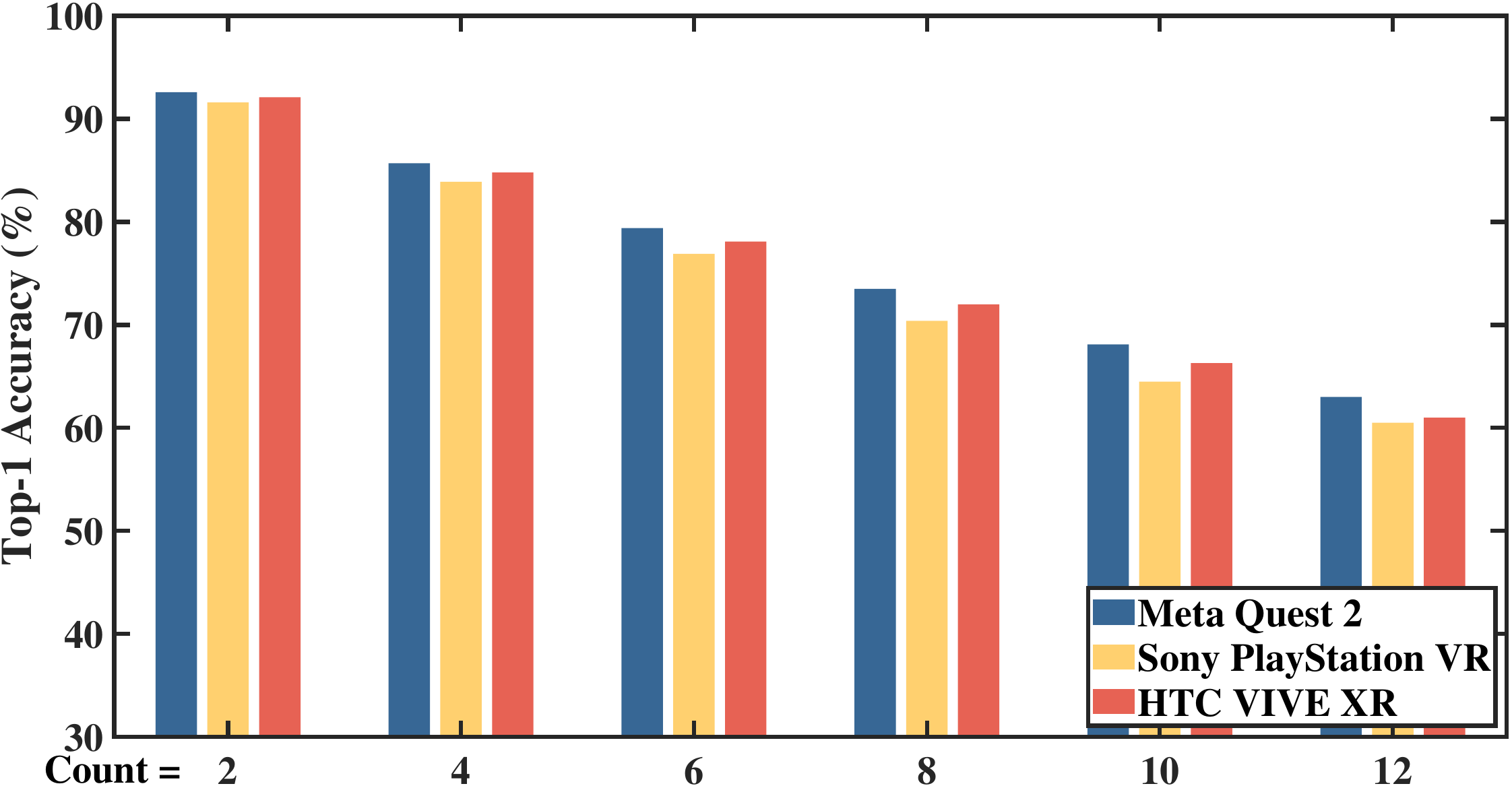}
        \caption{Top-1 Accuracy of private information recovery.}
        \label{fig:infotop1}
    \end{minipage}
    \hfill
    \begin{minipage}[c]{0.48\textwidth}
        \includegraphics[width=\linewidth]{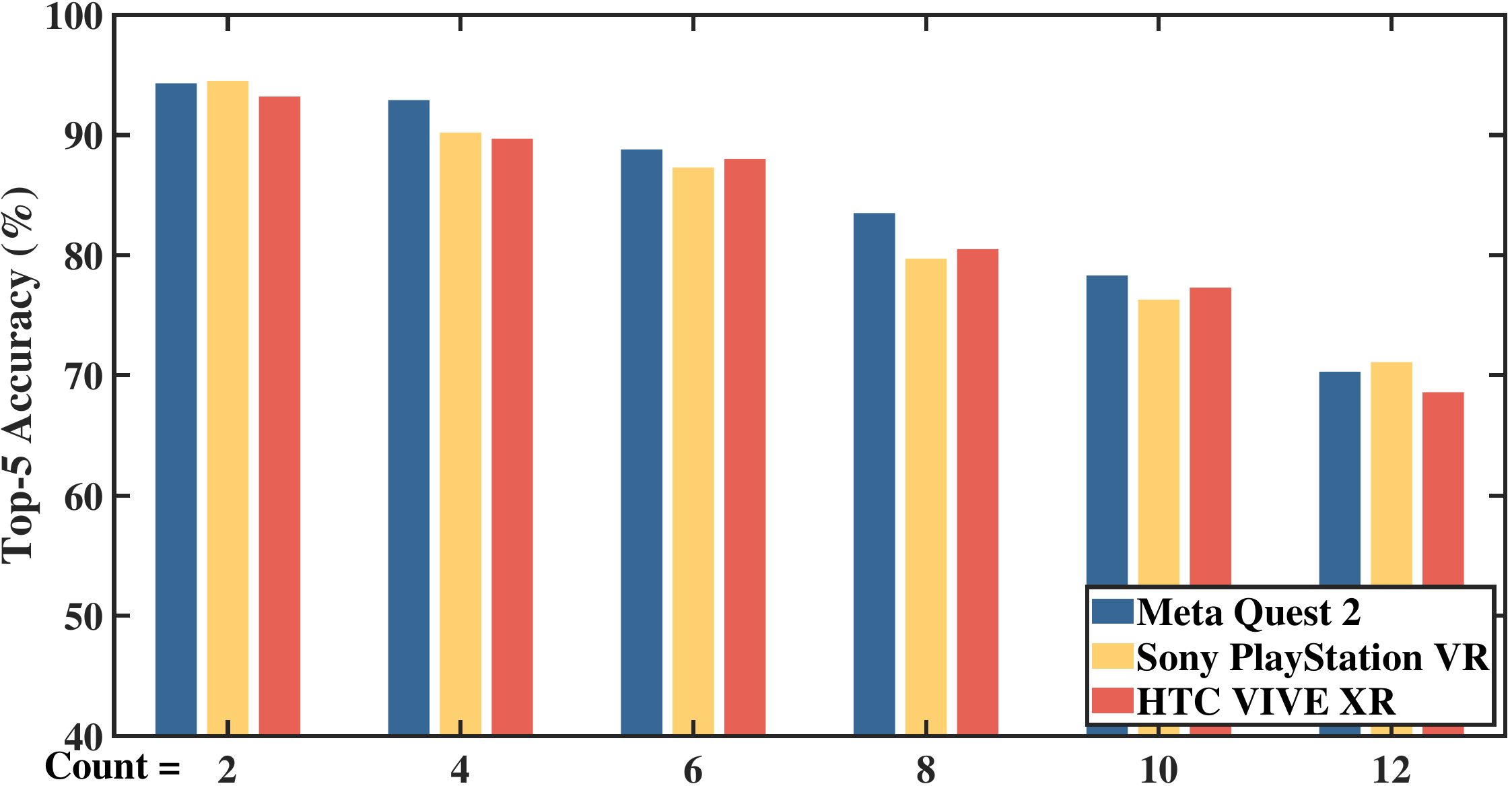}
        \caption{Top-5 Accuracy of private information recovery.}
        \label{fig:infotop5}
    \end{minipage}
\end{figure}

The evaluation of mmSpyVR's efficiency in recovering keystroke input is depicted in Figures~\ref{fig:infotop1} and~\ref{fig:infotop5}. This section assesses mmSpyVR's accuracy in recovering passwords entered by users on a 24-key VR keyboard. The "Count" in the figures represents the length of the password, indicating the number of characters input by the user, ranging from 2 to 12 characters. Given the critical nature of password privacy and its impact on user security, the security community considers an attack highly threatening if it achieves certain thresholds: single-character decryption accuracy exceeding about 85\%, overall password top-5 decryption accuracy surpassing about 60\%, and top-1 decryption accuracy greater than about 50\%~\cite{Privacy_Leakage1, Privacy_Leakage2, Privacy_Leakage3}. Therefore, we evaluate the top-1 and top-5 accuracy rates for different lengths of private information. 

The experimental results underscored the high accuracy of mmSpyVR in recovering private information, demonstrating robust performance across various password lengths and VR devices. We observe a gradual decline for both top-1 and top-5 accuracy metrics as password length increases from 2 to 12 characters. In the best-case scenario with 2-character passwords, mmSpyVR achieves an average top-1 accuracy of 92.1\% and a top-5 accuracy of 94.0\%. Even in the most challenging case of 12-character passwords, the system maintains an average top-1 accuracy of 61.5\% and a top-5 accuracy of 70.0\%. This trend reflects the cumulative effect of keystroke accuracy on overall precision in identifying longer sequences. Notably, mmSpyVR's performance remains consistent across different VR platforms, including Meta Quest 2, Sony PlayStation VR, and HTC VIVE XR, with only minor variations. These findings affirm mmSpyVR's capability to accurately reconstruct VR users' private information, showcasing its effectiveness and robustness even in complex conditions and across various VR platforms. The system's ability to maintain high accuracy rates, particularly for top-5 predictions on longer passwords, underscores its potential as a significant privacy threat in VR environments.

\begin{figure}[h]
    \centering
    \includegraphics[width=0.5\linewidth]{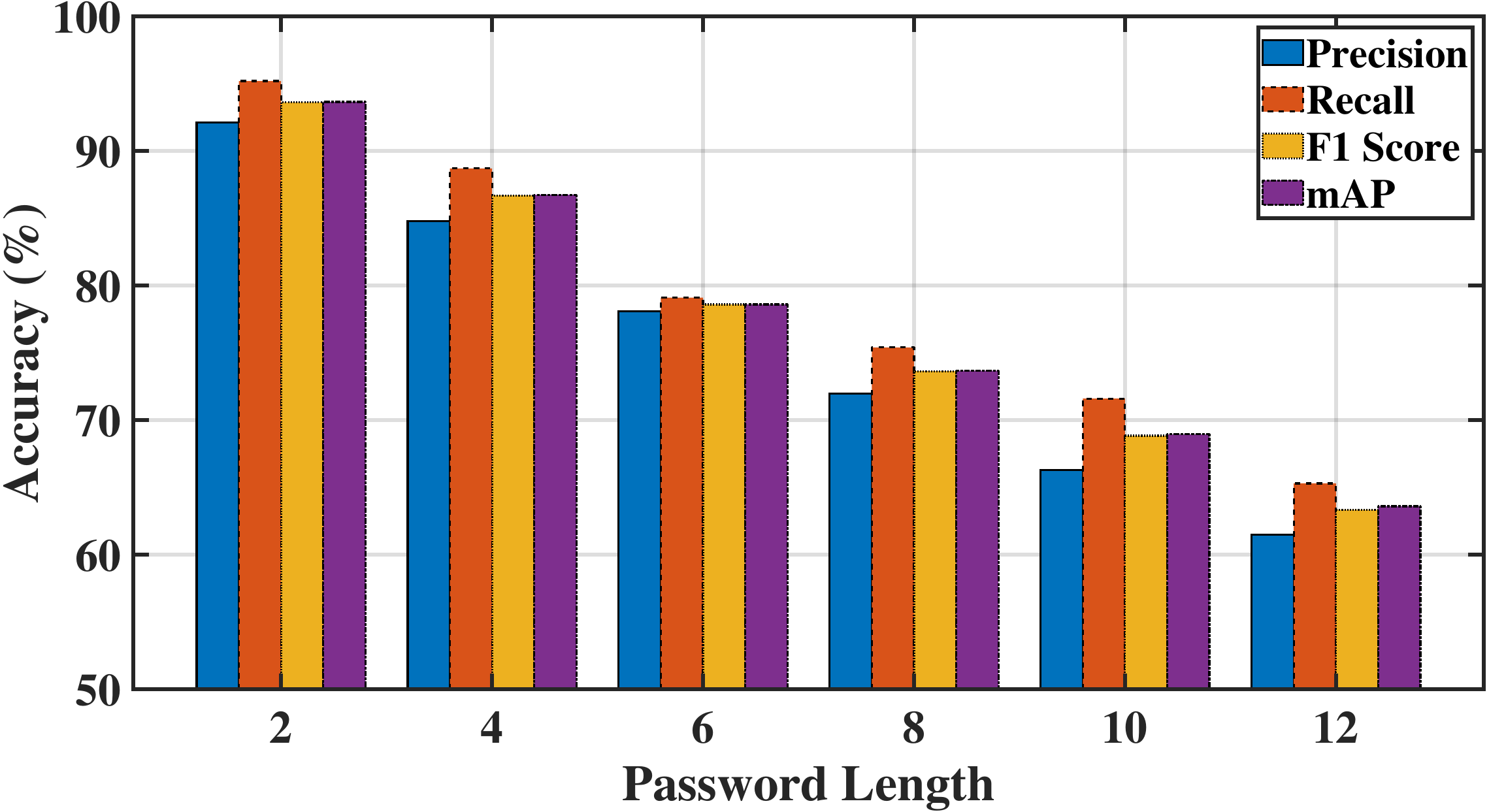}
    \caption{Statistical accuracy for different password lengths.}
    \label{fig:F1score}
\end{figure}

Fig.~\ref{fig:F1score} presents a detailed analysis of mmSpyVR's performance across various password lengths, ranging from 2 to 12 characters. The graph displays four key metrics: Precision, Recall, F1 Score, and mean Average Precision (mAP).  As shown in the figure, mmSpyVR demonstrates robust performance across the metrics, even as password length increases. For shorter passwords, the model achieves exceptionally high accuracy, with the metrics above 90\%. As expected, performance gradually declines as password length increases, but the model maintains impressive results even for longer passwords. Notably, the Recall metric consistently outperforms other metrics across various password lengths, indicating the model's ability to identify characters within passwords. The F1 Score, which balances Precision and Recall, remains high, demonstrating the model's overall effectiveness. The mAP metric, which considers performance across various threshold settings, also shows strong results, particularly for shorter passwords. Even for the most challenging 12-character passwords, mmSpyVR maintains a precision of 61.5\%, a Recall of 65.3\%, an F1 Score of 63.3\%, and an mAP of 63.6\%. These results underscore the model's capability to pose a significant threat to password security across various password complexities.
\section{Related Work}
\label{sec8_relatedwork}
Table~\ref{table:mmSpyVR} provides an overview of existing methods in motion tracking and VR spying, highlighting the novelty of our mmSpyVR approach. The mmSpyVR is the first to leverage mmWave technology for VR spying.

\textbf{Limitations of Non-mmWave VR Spying Methods.}
Existing VR spying techniques are categorized into three main approaches, each with significant drawbacks. Physical intrusion methods, such as hijacking internet cameras~\cite{Physiological}, face increasing challenges due to enhanced security measures and the risk of detection. These methods are also limited by obstacles and line-of-sight requirements~\cite{HoloLogger, Keylogging}, reducing their effectiveness in enclosed spaces. Virtual connection attacks, which compromise VR devices and establish virtual connections to infer user inputs~\cite{USENIX24, Face-Mic, FaceReader}, are becoming less viable due to improved device and network security, high detection risks, and inconsistencies between virtual and real-world hand motions. Wi-Fi-based attacks~\cite{Privacy_Leakage2, CSI:DeSpy, 3D_Human_Pose}, while less intrusive, struggle with limitations in bandwidth, carrier frequency, and signal robustness~\cite{Yan_2024_CVPR, GoPose}. These factors impair their ability to capture precise environmental profiles and resolve fine-grained VR user movements. Additionally, Wi-Fi signals are susceptible to interference from other environmental signals~\cite{Placement_Matters}, further compromising detection accuracy.

\begin{table}[h]
\centering
    \caption{Comparison of mmSpyVR and existing methods.}
    \begin{tabular}{|c|c|c|}
    \hline
     & \textbf{Motion Tracking} & \textbf{VR Spying} \\ \hline
    \textbf{Non-mmWave} & \cite{CSI:DeSpy, 3D_Human_Pose} & \cite{Privacy_Leakage1, Privacy_Leakage2, Privacy_Leakage3} \\ \hline
    \textbf{mmWave-based} & \cite{RadarNet, Tesla-Rapture, mmASL} & \textbf{mmSpyVR (Our Work)} \\ \hline
    \end{tabular}
\label{table:mmSpyVR}
\end{table}

In contrast, mmWave technology offers superior advantages for VR spying. mmWave signals are less affected by environmental interference, making them suitable for precise VR tracking. The high-frequency nature of mmWave enables highly accurate detection of hand positions, which is crucial for VR privacy spying and challenging for traditional low-frequency methods. Moreover, VR controllers and headsets are particularly effective at reflecting mmWave signals, enhancing the technology's suitability for this application. By leveraging these strong reflective properties, mmWave signals provide accurate and reliable data for VR privacy spying, overcoming the limitations of non-mmWave methods and offering a novel approach to this critical security concern.

\textbf{Existing mmWave-based Methodologies.}
While mmWave signals gain popularity in human sensing tasks due to their wide bandwidth and high carrier frequency~\cite{Multi_Modal_survey}, existing mmWave-based motion recognition techniques have limitations when applied to VR privacy spying~\cite{RadarNet, Tesla-Rapture}. The difference lies in the lack of direct one-to-one correspondence between VR user movements and their inputs. In VR environments, identical movements signify different inputs at different times, depending on the context and the sequence of actions. Consequently, even though existing mmWave sensing techniques achieve motion recognition~\cite{mmASL}, the sparse data obtained through walls is insufficient to extract fine-grained VR information~\cite{Wall_Matters} for VR spying.

To bridge this gap, our paper addresses several unique challenges associated with utilizing mmWave radar for VR privacy spying. These challenges include the randomness and diversity of VR motions, the limited VR privacy feature information available in VR motions, and the extraction of VR privacy information from the sparse mmWave signal reflections. To overcome these challenges, we develop novel approaches that bridge the gap between motion detection and VR privacy spying. Through comprehensive analysis and understanding of VR actions, we designed a data augmentation method based on VR action recognition to mitigate the impact of signal attenuation caused by obstacles. Furthermore, we created a contextual temporal feature understanding model capable of identifying VR privacy information embedded within VR actions. This model leverages the temporal sequence of VR actions to extract VR privacy information, going beyond simple motion recognition. By addressing these challenges, we present a novel side-channel vulnerability that exploits mmWave radar to penetrate obstacles to spy on VR users' privacy.
\section{Countermeasures}
\label{sec6_countermeasures}
Our study has shown the threat of mmWave VR privacy spying. To mitigate the threat, we identify several potential countermeasures to mitigate the privacy leakage issues associated with VR devices. These solutions are discussed and aim to provide clear future work and next steps to address the problem.

\textbf{Physical Barriers.} One effective method to prevent spying by attackers is to set up physical barriers that block mmWave signals. This is achieved by utilizing materials that absorb mmWave signals, making it challenging for attackers to capture useful data. However, this approach requires careful implementation to avoid interfering with the VR device's own functionality. The barriers should be designed to primarily block spying signals while allowing necessary communications between the VR device and the Internet access point. Additionally, consideration must be given to the practicality and aesthetics of such barriers in various VR use scenarios, from home environments to commercial VR arcades. While this method offers a robust defense against mmWave-based attacks, it should be balanced with user convenience and the overall VR experience.

\textbf{Software Obfuscation.} Another approach is to implement software-based obfuscation techniques. This approach involves dynamically altering the layout of the virtual keyboard in real-time, making it challenging for attackers to correctly interpret the VR user's actions. By continuously shuffling the positions of keys even changing the entire input method, the system creates a disconnect between the observed physical movements and their actual meaning in the virtual environment. For instance, the keyboard layout could randomly reorganize after each input, the system alternate between different keyboard types. This dynamic alteration of the keyboard layout disrupts the attacker’s ability to decipher the input, thereby enhancing the security of the VR environment. However, it's important to note that this method impacts the user experience and is, therefore, best employed in highly sensitive scenarios, such as when entering bank card PINs.

\textbf{Hardware Camouflage.} A third approach to enhance VR privacy involves hardware-based camouflage techniques. This method focuses on modifying the physical properties of VR devices to alter their interaction with mmWave signals. One strategy is to apply wave-absorbing materials to VR controllers and headsets. Alternatively, metamaterials are capable of being incorporated into the device design to manipulate the direction of mmWave signal reflections. By altering the reflection characteristics, VR devices could be disguised to the signal patterns of ordinary objects like furniture. This approach not only diminishes the quality and quantity of information available to eavesdroppers but also makes it challenging for attackers to distinguish VR devices from other innocuous objects in the environment, thereby enhancing the overall security of VR systems. While this approach requires some modifications to existing VR hardware designs, it offers a passive and consistently active form of protection against mmWave-based privacy attacks.

\textbf{Environmental Monitoring.} Moreover, we propose leveraging the existing mmWave radar technology present in some VR devices to enhance security through active environmental monitoring. These built-in radars are able to be repurposed to scan the surrounding area for potential eavesdroppers and suspicious signal activities. The VR system analyzes the radar data to identify unusual patterns and anomalies that might indicate the presence of an attacker. This proactive approach would enable threat detection, allowing the system to alert users immediately when a potential security risk is identified. For instance, if an unexpected signal reflection pattern is detected, suggesting the presence of an unauthorized spying device, the system notifies the user and potentially triggers additional security measures, e.g., changing the keyboard layout. This method not only adds an extra layer of security but also empowers users with increased awareness of their digital surroundings, making it challenging for attackers to operate undetected in VR environments.
\section{Reporting and Discussion}
\label{sec7_discussion}

We reveal the privacy leakage problem of VR devices exploited by such an attack. To validate our findings, we report them to Meta~\cite{Meta}, a leading VR company, and ask for their feedback.

\begin{table}[h]
\centering
\caption{Privacy Risk Quantification Matrix}
\label{tab:privacy}
\resizebox{0.9\linewidth}{!}{
\begin{tabular}{lll}
\toprule
    \textbf{Leaked Privacy Content} & \textbf{Privacy-Leaking Actions} & \textbf{Privacy Scale} \\
    \midrule
    Banking Credentials & Hand and head movements during shopping & Very High \\
    Wi-Fi Passwords &  Hand and head movements during system setting & Very High \\
    Social Media Logins & Hand and head movements during chatting & High \\
    Private Messages & Hand and head movements during messaging & High \\
    Game Account Details & Hand movements during game login and in-game purchases & High \\
    Other Keyboard Input & Hand and head movements during virtual keyboard interaction & High \\
    User Habits & Patterns in body movements and activity types & Medium-High \\
    Browsing Preferences & Body movements and hand gestures during web browsing & Medium-High \\
    Shopping Preferences & Body movements during virtual shopping experiences & Medium \\
    Other Activity Type & Body movements, intensity and frequency of actions & Medium \\
    \bottomrule
    \end{tabular}
}
\end{table}

\subsection{Privacy Leakage in VR Devices}
VR devices offer users a variety of activities that enrich their daily lives. However, they also pose significant privacy threats. Existing researches have shown that sensitive data are leaked through users' motions during VR interactions~\cite{Privacy_Leakage1,Privacy_Leakage2,Privacy_Leakage3}. To illustrate the breakdown of privacy leakage in VR, we present Table~\ref{tab:privacy}, which categorizes potential privacy breaches based on two key dimensions: Privacy-Leaking Actions and Privacy Scale~\cite{Privacy_risk_models}. The vulnerabilities involve critical information~\cite{10.1145/3503161.3548386, 9382914} such as the type of ongoing activity. Furthermore, attackers may reveal privacy information entered by VR users~\cite{De-anonymization}, including Wi-Fi passwords, bank and social account credentials~\cite{Valluripally_Modeling, Valluripally_Detection}, private message content~\cite{Radio2Text}, and click patterns.

Specifically, privacy leakage in VR devices occurs when attackers exploit users' motions to identify their activities and keystroke inputs. These motions are divided into body and hand movements. Body movements, which vary in intensity and frequency, help distinguish between different types of activities. For example, gaming often involves dynamic actions like jumping and crouching, while chatting apps are characterized by gestures like waving~\cite{ZGaming}. Hand movements provide a detailed view of the user's interactions, especially in activities requiring text and number input, such as chatting, shopping, and browsing, where users interact with a virtual keyboard~\cite{VRChat23}. These interactions reveal the timing and position of finger taps, allowing attackers to reconstruct keystrokes and potentially access sensitive data, like passwords and account information. This exploitation of detailed motion data underscores the significant privacy risks inherent in VR.

\subsection{Threat Model Justification} 
We consider a typical attack scenario in this line of related research (e.g., ~\cite{HiddenReality}), where a victim plays a VR system in a private area (e.g., home and office shown in Fig.~\ref{fig:scenario}). Attackers place an attacking device behind a brick wall and a wooden door that is capable of transmitting mmWave to penetrate building materials and capture the reflection by the VR user. As such, the VR user is unaware of the presence of the attacking device because it is out of sight~\cite{9637169}. Additionally, we assume that the adversary \textit{cannot} compromise both the hardware (e.g., the headset and controllers) and the software (e.g., OS) of the VR system.

The threat model presented in this section is realistic and feasible, as it is based on the existing capabilities and limitations of mmWave technology and VR devices. The mmWave signals penetrate various non-metallic materials, such as wood, brick, plastic, and glass, with minimal attenuation~\cite{mmPhone, MILLIEAR}. Moreover, mmWave signals reflect and scatter from user bodies, capturing fine-grained information about their movements and gestures. Therefore, an attacker utilizes an mmWave device to launch attacks on a VR user without a physical and virtual connection with VR users. We place the mmWave device outside the VR user’s room so that the VR user move freely inside the room without aware of the mmWave device’s presence. The distance between the VR user and the mmWave device varies from 2 to 8 meters as the VR user moves inside the room.

\subsection{Industry Response and Future Directions}
To summarize, Meta confirms that VR privacy spying on VR devices utilizing mmWave technology is feasible and that such technology reveals private user information. They also acknowledge that VR devices are widely utilized for gaming~\cite{ZGaming}, and that many games require logging in with Google and Facebook accounts~\cite{AnalysisReview}, which poses a serious privacy risk. Furthermore, they recognize that VR devices are increasingly utilized for work-related purposes~\cite{Non-Contact, Physiological, VRChat23}, such as video meetings~\cite{ThingShare} and chatting~\cite{RemoteTouch}, especially with the launch of the Apple Vision Pro~\cite{Apple}, which makes the potential impact of mmSpyVR’s exploitation of vulnerabilities severe. Applying mmSpyVR to detect this issue demonstrates the effectiveness and novelty of our attack and urges manufacturers to address such risks before they cause damage.

Moreover, we explore in detail the feasibility and potential implementation of various countermeasures, including physical barriers, software obfuscation, hardware camouflage, and environmental monitoring, as discussed in the countermeasures section. Meta is interested in these approaches and recognizes the importance of enhancing privacy in VR environments. They are receptive to incorporating some of our proposed solutions into future product designs. Furthermore, Meta is willing to collaborate with academic researchers on future work. This positive response validates the significance of our findings and paves the way for potential joint efforts to develop and implement effective privacy protection measures in next-generation VR devices.
\section*{ACKNOWLEDGMENTS}
We sincerely thank the anonymous area chair and reviewers for their valuable comments. This work was supported in part by the National Natural Science Foundation of China under Grant No. 62272098 and the Ministry of Education, Singapore, under its Joint SMU-SUTD Grant (22-SIS-SMU-052) and NSF China 62102332, ECS CityU 21216822, and City University of Hong Kong 9610491.

\newpage
%
\bibliographystyle{ACM-Reference-Format}
\bibliography{sample-base}

\section{IRB Compliance Declaration}
\label{sec:irb_compliance}
In our latest Progress Report, we documented our adherence to the highest ethical standards as mandated by our Institutional Review Board (IRB):
    
\begin{itemize}
    \item \textbf{Participant Recruitment}: We employed convenience sampling through university mailing lists and local community boards. Our selection criteria included factors such as prior VR experience and absence of health conditions that could be affected by mmWave exposure.
    
    \item \textbf{Informed Consent}: All participants were provided with comprehensive information about the study, including potential health risks associated with mmWave trials. Our IRB-approved consent form explicitly outlined these risks and the measures taken to mitigate them, ensuring full transparency.
    
    \item \textbf{Ethical Review Tracking Frequency}: We have established a rigorous schedule for ethical review tracking, which is conducted at regular intervals (annually, biannually, quarterly, or as specified). 
        
    \item \textbf{Study Situations and Risks}: We confirm that no circumstances adversely impact study progress. Furthermore, there have been no instances where study risks have surpassed anticipated levels. 
    
    \item \textbf{Safety Measures}: Throughout the study, we adhered to strict safety protocols. Participants were regularly monitored for any discomfort, and exposure to mmWave radiation was kept within established safety limits as per regulatory guidelines. 
\end{itemize}

This declaration affirms that our research is conducted without ethical breaches and within the scope of IRB-approved protocols. We remain committed to upholding the standards of research ethics and integrity.

\end{document}